\documentclass[fleqn,usenatbib]{rasti}

\usepackage{newtxtext,newtxmath}

\usepackage[T1]{fontenc}

\usepackage[table]{xcolor}

\DeclareRobustCommand{\VAN}[3]{#2}
\let\VANthebibliography\thebibliography
\def\thebibliography{\DeclareRobustCommand{\VAN}[3]{##3}\VANthebibliography}

\usepackage{graphicx}	
\usepackage{amsmath}	
\usepackage{enumitem}  

\usepackage{pdflscape}	

\title[Synergies between dust and heliosphere science]{Synergies between interstellar dust and heliospheric science with an Interstellar Probe}

\author[V. J. Sterken et al.]{Veerle J. Sterken,$^{1}$\thanks{E-mail: vsterken@ethz.ch (VJS)}
S. Hunziker,$^{1}$ 
K. Dialynas,$^{2}$ 
J. Leitner,$^{3}$ 
M. Sommer,$^{4}$ 
R. Srama,$^{4}$ 
L. R. Baalmann,$^{1}$ \newauthor
A. Li,$^{5}$
K. Herbst,$^{6}$  
A. Galli,$^{7}$
P. Brandt,$^{8}$ 
M. Riebe,$^{9}$
W. J. Baggaley,$^{10}$
M. Blanc,$^{11}$  
A. Czechowski,$^{12}$  \newauthor 
F. Effenberger,$^{13,14}$      
B. Fields,$^{15}$
P. Frisch,$^{16}$
M. Horanyi,$^{17}$ 
H.-W. Hsu,$^{17}$   
N. Khawaja,$^{18}$   
H. Kr\"{u}ger,$^{19}$ \newauthor
W. S. Kurth,$^{20}$
N.F.W. Ligterink,$^{7}$
J. L. Linsky,$^{21}$       
C. Lisse,$^{8}$
D. Malaspina,$^{17,22}$ 
J.A. Miller,$^{23}$ 
M. Opher,$^{23}$ \newauthor
A. R. Poppe,$^{24}$
F. Postberg,$^{18}$      
E. Provornikova,$^{8}$
S. Redfield,$^{25}$
J. Richardson,$^{26}$   
M. Rowan-Robinson,$^{27}$
K. Scherer,$^{13}$   \newauthor
M. M. Shen,$^{28}$
J. D. Slavin,$^{29}$    
Z. Sternovsky,$^{17}$ 
G. Stober,$^{30}$
P. Strub,$^{4,19}$    
J. Szalay,$^{28}$ 
and M. Trieloff$^{3}$
\\
$^{1}$Institute for Particle Physics and Astrophysics, Physics Dept., ETH Z\"{u}rich, Wolfgang-Pauli Strasse 27, Z\"{u}rich, Switzerland.\\
$^{2}$Center of Space Research and Technology, Academy of Athens, 4, Soranou Efesiou str., 11527, Papagos, Athens, Greece.\\
$^{3}$Institute of Earth Sciences, Ruprecht-Karls-Universität Heidelberg, Im Neuenheimer Feld 234-236, 69120 Heidelberg, Germany.\\
$^{4}$Institute Space Systems (IRS), University Stuttgart, Pfaffenwaldring 29, 69168, Stuttgart, Germany.\\
$^{5}$Department of Physics and Astronomy, University of Missouri, Columbia, MO 65211, USA.\\
$^{6}$Inst. of Exp. and Applied Physics (IEAP), Fac. of Math. and Natural Sciences, Christian-Albrechts-University of Kiel, Leibnizstrasse 11-19, 24118, Kiel, Germany.\\
$^{7}$Physics Institute, University of Bern, 3012, Bern, Switzerland.  \\
$^{8}$Applied Physics Laboratory, The Johns Hopkins University, Laurel, Maryland 20723, USA.  \\
$^{9}$Institute of Geochemistry and Petrology, Dept. of Earth Sciences, ETH Z\"{u}rich, Clausiusstrasse 25, 8092 Zürich, Switzerland. \\
$^{10}$Department of Physics and Astronomy, University of Canterbury, Christchurch, New Zealand.\\
$^{11}$Institut de Recherche en Astrophysique et Planétologie (IRAP), 9 Av. du Colonel Roche, 31400 Toulouse, France. \\
$^{12}$Space Research Centre, Polish Academy of Sciences, Warsaw, Poland. \\
$^{13}$Institute for Theoretical Physics IV, Chair for Space- and Astrophysics, Ruhr-Universit\"{a}t Bochum, D-44780 Bochum, Germany.\\
$^{14}$Bay Area Environmental Research Institute, NASA Research Park, Moffett Field, CA, USA.  \\
$^{15}$Center for Theoretical Astrophysics, Departments of Astronomy and Physics, University of Illinois, Urbana, IL 61801, Illinois, USA.  \\
$^{16}$University of Chicago, Astronomy \& Astrophysics Center, 5640 S Ellis Ave, Chicago, 60637, Illinois (IL), USA. \\
$^{17}$Laboratory for Atmospheric and Space Physics, University of Colorado, Boulder, CO, USA \\
$^{18}$Institute of Geological Sciences, Freie Universit\"{a}t Berlin, D-12249 Berlin, Germany. \\
$^{19}$Max-Planck-Institut f\"{u}r Sonnensystemforschung, Justus-von-Liebig-Weg 3, 37077 G\"{o}ttingen, Germany.\\
$^{20}$Department of Physics and Astronomy, University of Iowa, Iowa City, Iowa, USA.  \\
$^{21}$JILA, University of Colorado, Boulder 80305-0440, Colorado, USA. \\
$^{22}$Astrophysical \& Planetary Sciences Department, University of Colorado Boulder, CO, USA.  \\
$^{23}$Astronomy Department, Boston University, Cambridge, Massachusetts, USA. \\
$^{24}$Space Sciences Laboratory, University of California at Berkeley, Berkeley, CA 94720, USA.  \\
$^{25}$Astronomy Department, Van Vleck Observatory 101, Wesleyan University, 96 Foss Hill Drive, Middletown, CT 06459, Connecticut, USA.  \\
$^{26}$ MIT Kavli Inst. for Astroph. and Sp. Res., 77 Massachusetts Avenue, McNair Building (MIT Building 37) MA 02139, Cambridge, Massachusetts, USA.  \\
$^{27}$Astrophysics Group, Imperial College London, Prince Consort Rd, London SW7 2BZ, United Kingdom. \\
$^{28}$Princeton University, Princeton, NJ 08544, USA. \\
$^{29}$Center for Astrophysics, High Energy Astrophysics Division, Harvard \& Smithsonian, 60 Garden Street, MS 04, Cambridge 02138-1516, Massachusetts, USA.  \\
$^{30}$Microwave Physics, Institute of Applied Physics \& Oeschger Center for Climate Change Research, University of Bern, Bern, Switzerland. \\
}

\date{Accepted 01/08/2023. Received 02/07/2023; in original form 07/12/2022 \\ \\ This article has been accepted for publication in RASTI Published by Oxford University Press on behalf of the Royal Astronomical Society.}

\pubyear{2023}

\begin{document}
\label{firstpage}
\pagerange{\pageref{firstpage}--\pageref{lastpage}}
\maketitle

\begin{abstract}
We discuss the synergies between heliospheric and dust science, the open science questions, the technological endeavors and programmatic aspects that are important to maintain or develop in the decade to come. In particular, we illustrate how we can use interstellar dust in the solar system as a tracer for the (dynamic) heliosphere properties, and emphasize the fairly unexplored, but potentially important science question of the role of cosmic dust in heliospheric and astrospheric physics. We show that an Interstellar Probe mission with a dedicated dust suite would bring unprecedented advances to interstellar dust research, and can also contribute –- through measuring dust -– to heliospheric science. This can, in particular, be done well if we work in synergy with other missions inside the solar system, thereby using multiple vantage points in space to measure the dust as it ‘rolls’ into the heliosphere. Such synergies between missions inside the solar system and far out are crucial for disentangling the spatially and temporally varying dust flow. Finally, we highlight the relevant instrumentation and its suitability for contributing to finding answers to the research questions. 
\end{abstract}




\section{Introduction and background information}
This paper discusses the synergies\footnote{A Synergy: the interaction or cooperation of two or more organizations, substances, or other agents to produce a combined effect greater than the sum of their separate effects. [Oxford Languages dictionary]} between heliospheric and dust science that can be harnessed with an interstellar probe, the open science questions and pathways forward in the future, including the relevant instrumentation. We refer to \cite{sterken:2019} for a review of the current state of the art of interstellar dust research in the solar system (dynamics and composition, measurements and models). This paper (RASTI) was originally submitted to the Decadal Survey for Solar and Space Physics (Heliophysics) 2024 \--- 2033 – 7 Sept. 2022 and would be published in the Bulletin of the AAS (BAAS):~\citet{sterken:2022} \--- doi pending. It is modified in this version and includes a discussion on the instrumentation necessary to answer the science questions. Two accompanying white papers were submitted for the decadal survey: \cite{hsu:2022}, ``Science opportunities enabled by in situ cosmic dust detection technology for heliophysics and beyond'', and \cite{poppe:2022}, ``The interactions of Interstellar Dust with our Heliosphere''. A third accompanying refereed paper~\citep[][in prep.]{hunziker:2023isp}, will provide dust flux predictions in order to illustrate how dust measurements on the way out through the heliosphere may provide new constraints (i.e., the boundary conditions) for heliosphere models, in addition to the already existing magnetic field, plasma, Galactic Cosmic Ray (GCR) and other data from the Voyagers and other spacecraft. 

\subsection{The solar system in the Local Interstellar Cloud}\label{sec:ss_in_lic}
The Sun and planets move through the outer edges of the local interstellar cloud (LIC) and into the neighbouring G-cloud or a mixed region of the two clouds \citep{swaczyna:2022} after a journey of nearly 60.000 years in the LIC \citep{linsky:2022}. The interstellar dust (ISD) in this diffuse cloud may have its origins in supernovae and atmospheres of cool stars or may be recondensed in the interstellar medium after being shattered by supernova shocks. These particles cross the solar system due to its relative velocity with the LIC (of about 26 km\,s$^{-1}$). They can be measured in situ by dust detectors on spacecraft, and hereby provide unique ground truth information about their make-up and dynamics. This ground truth information is complementary to measurements of the dust by more classical astronomical methods like observations of extinction, scattering, and polarisation of starlight as well as dust thermal emission, and by observing the gas in comparison to a reference (the so-called “cosmic abundances”, usually the solar composition), where the “missing component” in the gas phase hints at what must be locked up in the dust \citep{mathis:1977, draine:2003, Draine2007, draine:2009, wang:2015}. Directly measuring these particles is of utmost importance for astrophysics and is also part of humanity’s exploration of our local interstellar neighbourhood.

\begin{figure}
	\includegraphics[width=\columnwidth]{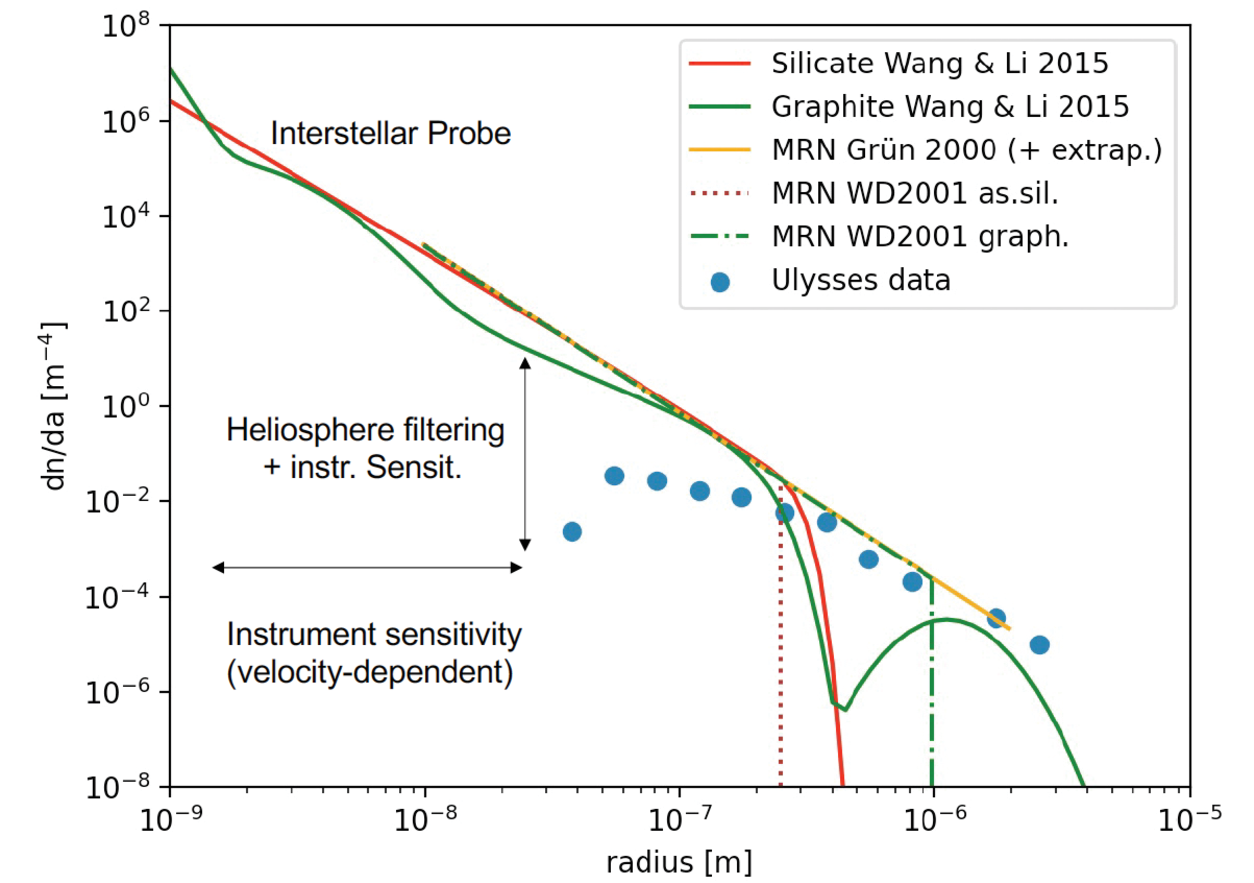}
    \caption{The interstellar dust size distribution from astronomical models and in situ data by Ulysses \citep[from][]{sterken:2022}. The smallest interstellar dust particles are the most numerous. Distributions derived from astronomical observations were taken from~\citet{wang:2015, gruen:2000, weingartner:2001}. Ulysses data are from~\citet{krueger:2015}.}
    \label{fig:ISD_size_distr}
\end{figure}
\begin{figure}
    \includegraphics[width=\columnwidth]{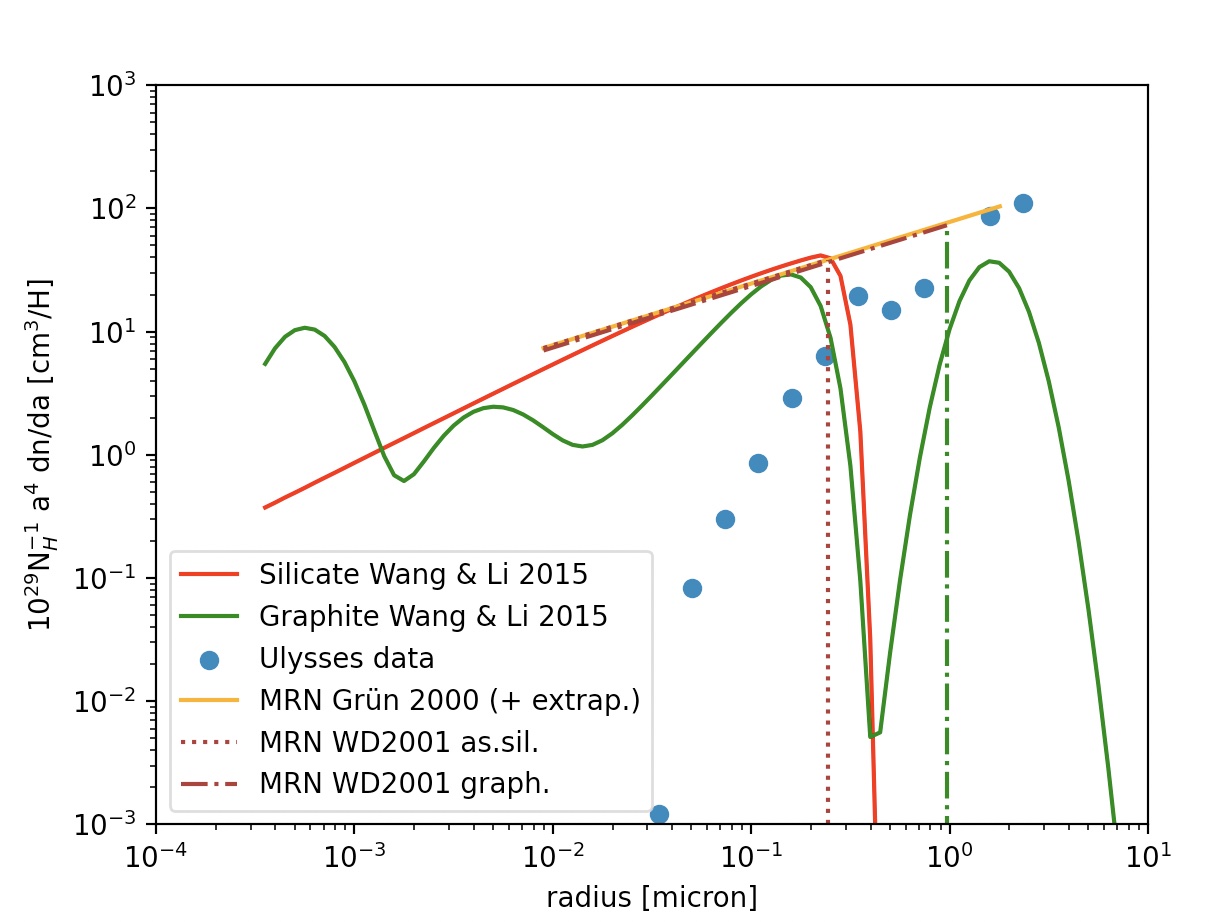}
    \caption{The interstellar dust mass distribution from astronomical models and in situ data by Ulysses. Most of the mass is in the largest particles, which is important for the determination of the gas-to-dust mass ratio. Units on the vertical axis are chosen to be equal to the units in~\citet{wang:2015}. Distributions derived from astronomical observations were taken from~\citet{wang:2015, gruen:2000, weingartner:2001}. Ulysses data are from~\citet{krueger:2015}.} 
    \label{fig:ISD_mass_distr}
\end{figure}

\subsection{Dynamics of interstellar dust in the heliosphere}
\begin{figure}
	\includegraphics[width=\columnwidth]{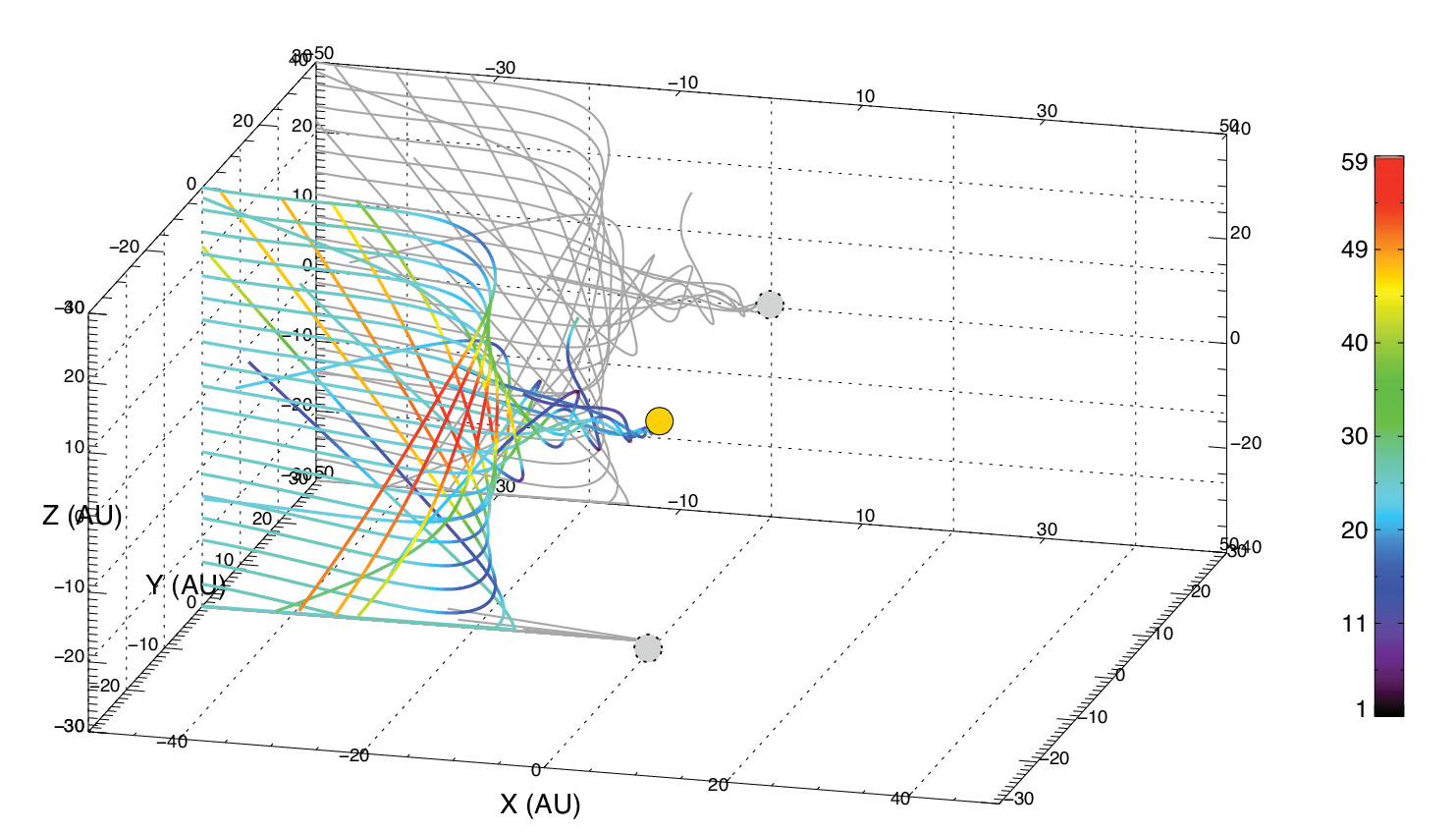}
    \caption{A complicated pattern of ISD trajectories for charge-to-mass ratio 12~C/kg, assuming they made it through the heliosheath \citep{sterken:2012}.}
    \label{fig:complicated_pattern}
\end{figure}

\begin{figure}
	\includegraphics[width=\columnwidth]{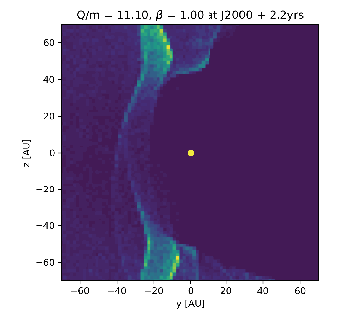}
    \caption{Higher dust densities `rolling' into the heliosphere for Q/m = 11.1~C/kg \citep{hunziker:2023isp}.}
    \label{fig:rolling_waves}
\end{figure}

The interstellar dust (ISD) size distribution extends from nanometers to several micrometers and decreases with increasing particle size (Figure~\ref{fig:ISD_size_distr}). However, its mass distribution increases with particle size, as illustrated in Fig.~\ref{fig:ISD_mass_distr} \citep[see also, for example,][Fig.~6]{krueger:2015}, and thus the largest ISD particles are the most important for determining the gas-to-dust mass ratio (R$_{\rm g/d}$) in the LIC. The dynamics of the dust in the heliosphere depends on the particle size, optical properties, and on the space environment. This dependence on the space environment turns ISD into a very interesting tracer for the dynamic heliosphere. \\
\\
{\bf Micron-sized ISD} particles passing through the solar system are gravitationally dominant, may be uncoupled from the LIC, and could, in theory, come from any other direction than the heliosphere nose (note that the interstellar meteoroids are still a controversial topic in the field, e.g.,~\citet{hajdukova:2020, hajdukova:2023, brown:2023}).\\ 
\\
{\bf Mid-sized ISD} particles (ca. 0.1 \--- 0.6 $\mu$m radius) can reach the solar system depending on their size, optical properties, composition, and phase in the solar cycle. Their dynamics in the solar system are governed by solar gravitation, by solar radiation pressure force, and by Lorentz forces due to (charged) ISD passing through the magnetic fields of the solar wind plasma that changes with the 22-year solar cycle, leading to an alternating focussing and defocussing of the dust towards the solar equatorial plane during the solar minima. However, there is an additional (most likely time-dependent) mechanism of filtering in the heliosheath \citep{linde:2000, slavin:2012, sterken:2015}.\\
\\
{\bf Small ISD} particles (30-100 nm) are dominated by the Lorentz force and may partially reach the solar system during the solar focussing phase (e.g., 2029-2036) if the heliospheric boundary regions do not filter the particles already upfront. The higher the charge-to-mass ratio of the dust is, the more the particles move on complicated patterns (e.g., Figure~\ref{fig:complicated_pattern}, from \cite{sterken:2012}), which may cause ‘waves’ of higher dust densities to ‘roll’ into the heliosphere for specific particle sizes (Figure~\ref{fig:rolling_waves}, from \cite{hunziker:2023isp}). The exact lower cut-off size and time-dependence of particles that can enter the solar system is not yet exactly known, but Ulysses and Cassini already have measured ISD particles with radii between 50 and 100~nm~\citep{altobelli:2016, krueger:2015}.\\ 
\\
{\bf Nanodust} (2\---30 nm) cannot enter the heliosphere because it is coupled to the magnetic field lines of the very local interstellar medium (VLISM), and is diverted around the heliopause boundary \citep{linde:2000, slavin:2012}. These particles may also pile-up at the heliopause \citep{slavin:2012, frisch:2022}. Polycyclic aromatic hydrocarbon (PAH) molecules are the smallest carbon nanodust particles in the interstellar medium. They are abundant and widespread in a wide variety of astrophysical regions \citep{Li:2020}. Their presence (or absence) in the local interstellar cloud would provide useful insight into the nature and origin of interstellar PAHs. They are not expected to enter the solar system since they would have been deflected from the heliosphere. However, if PAHs of interstellar origin {\it are} detected in the solar system or beyond, their origin (possibly through fragmentation of small carbon dust) would offer valuable insight into the composition and structure of interstellar carbon dust.

\subsection{Filtering of dust in the outer and inner boundary regions of the heliosphere}
The filtering of interstellar dust in the heliosphere likely happens mainly at the heliosphere boundary regions (i.e., at the heliopause and in the heliosheath) and in the region closer to the Sun because (1) the dust acquires the highest charges in the heliosheath \citep{kimura:1998, slavin:2012}, and (2) the azimuthal component of the interplanetary magnetic field causing the focusing and defocusing effects is the largest closer to the Sun. Flying a spacecraft through \underline{all these regions} to measure all parameters simultaneously (magnetic field, plasma densities, dust charging, pickup ions, dust flux, velocity, and direction), would be of utmost value for understanding the mechanisms of the dynamics of the dust, the dust-plasma interaction, and the role of dust for heliosphere physics, in particular, because this has never been done before. We, therefore, need in-situ measurements of dust and plasma parameters in interplanetary space, at the termination shock, in the heliosheath, at the heliopause and – especially – beyond the heliopause, over the solar cycle, from a future Interstellar Probe. \\
\\
In situ ISD measurements with Ulysses up to 5.4 AU indeed contained a signature of the dynamic heliosphere \citep{landgraf:2003, strub:2015, sterken:2015}. These particles were measured for the first time in 1993, using an impact ionization dust detector \citep{gruen:1993}. Ulysses monitored the ISD throughout the solar cycle for 16 years, giving us an impression of the fluxes and roughly the flow directionality of the dust. The dust flow direction changed, in particular, in the latitudinal direction around 2006 \citep{krueger:2007}, which may have been caused by the Lorentz force \citep{sterken:2015}. Simulations of ISD dynamics in the solar system (without heliospheric boundaries) would be piecewise compatible with the data if the larger dust particles are porous, or aggregates (hence, if they have a higher charge-to-mass ratio, \citet{sterken:2015}). A second time-dependent mechanism of filtering in the heliosheath was suggested to be needed in order to explain the Ulysses data \citep{sterken:2015}. Time-dependent models of the heliosphere-dust interaction including a heliosheath are currently under development. 

\subsection{Compositional measurements of interstellar and interplanetary dust grains}
\label{sec:composition}
\subsubsection{Circumstellar grains versus ISM dust versus solar system dust}
Both model predictions~\citep{zhukovska:2016} and analysis of presolar dust grains from primitive meteorites~\citep{hoppe:2017} indicate that only about $\sim$3~$\%$ of the interstellar medium (ISM) and parent interstellar cloud dust of our Sun is circumstellar dust. Measurements of contemporary interstellar dust offer the opportunity to compare the inferred ratios at the time of formation of the Solar System with present-day data. However, this would require a sufficiently large number of investigated particles. Assuming  a similar circumstellar dust/ISM dust ratio as obtained from modelling and at the time of Solar System formation, we would expect one circumstellar grain among $\sim$33~ISM grains, on average.\\
\\
From an astrophysical point of view, oxygen isotopes are best-suited to identify circumstellar oxygen-rich grains (silicates and refractory oxides), while carbon isotopes are diagnostic for most C-rich species~\citep{zinner:2014}. The oxygen and carbon isotopic compositions of the local ISM are not well constrained; there is information on the $^{18}$O/$^{17}$O  ratio~\citep[e.g., ][]{wouterloot:2008}, which differs from the Solar System value. Analyzing the the oxygen and carbon isotopic compositions of a large number of interstellar dust (ISD) particles would help closing this knowledge gap.\\
\\
Laboratory isotope measurements of extraterrestrial samples with high-spatial resolution Secondary Ion Mass Spectrometry (SIMS), like NanoSIMS~\citep{hoppe:2013} or TOF-SIMS~\citep{stephan:2001}, have to deal with a multitude of mass interferences, which often require high mass resolutions m/$\Delta$m of several thousand to be resolved. Spacecraft-mounted impact-ionization TOF-MS instruments, on the other hand, achieve mass resolutions $\le$200, which is sufficient to resolve different isotopes of C, O, Mg, Al, Si, and Ca but not sufficient to resolve the isotopes from interfering compound ions.  Nevertheless, due to the different ionization process, many compound ions responsible for hydride or oxide mass interferences (e.g., $^{24}$Mg$^{1}$H$^{+}$, $^{28}$Si$^{16}$O$^{+}$) do not occur in relevant numbers, allowing the measurement of diagnostic isotopic ratios, when the detection sensitivity of the element in question is sufficient. Limiting factors would be the concentration of the respective element in a given dust grain, and the impact velocity/energy, governing the ionization yield. The required impact velocities for different species are listed in Table~\ref{tab:velocities_comp}.\\ 
\\
Besides oxygen and carbon (which are detected as O– and C–), further elements of interest would be Mg, Al, Si, and perhaps Ca, all forming positive ions, and present in the vast majority of O-rich circumstellar grains (silicates and refractory oxides). Mg and Si in circumstellar grains can also display isotopic anomalies, although not as pronounced as O and C. However, the Mg- , Si-, and Ca-isotopic compositions of interstellar dust are unknown, making such measurements even more valuable. Another electronegative element of interest would be Sulfur, which is present in C-rich presolar grains~\citep{hoppe:2012}. Sulfides have been identified around evolved stars~\citep{hony:2002}. No presolar/circumstellar sulfides have been unambiguously identified so far, except for one signature in an impact crater on Stardust Al foils,~\citet{heck:2012}. None of the Cassini in situ measurements of 36 ISD particles~\citep{altobelli:2016} show evidence for sulfides, despite evidence that S is depleted from the gas phase in the ISM where an abundance of $\sim$100 ppm has been inferred for primitive Solar System materials~\citep{keller:2011}. Thus, we would expect a certain amount of circumstellar sulfides in the interstellar dust population, if sulfides are able to escape destruction in the ISM – an important question that could be addressed by in situ dust measurements. The mass of S coincides with that of O$_{2}$ and sulfur measurements, therefore, require a higher mass resolution than 200.

\subsubsection{Galactic Chemical Evolution}
Elemental – and especially isotopic – ratios of contemporary circumstellar and interstellar dust would greatly complement and enhance our knowledge on Galactic Chemical Evolution (GCE), i.e., the enrichment of the ISM and stars with heavier elements and heavier isotopic compositions over time. For certain elements, like Si and Mg, GCE-related correlations have been observed in presolar dust grains~\citep[e.g.,][]{zinner:2006, hoppe:2021}, and general predictions have been made for these and other elements from model calculations~\citep[e.g.,]{timmes:1995, kobayashi:2020}. However, models and measured grain data do not always show good correlations, thus, information on isotopic ratios like $^{25}$Mg/$^{24}$Mg, $^{26}$Mg/$^{24}$Mg, $^{29}$Si/$^{28}$Si and $^{44}$Ca/$^{40}$Ca would establish another baseline and allow the study of potential heterogeneities of these isotopic systems in the local ISM, which is typically not covered by the models. Similarly, $^{13}$C/$^{12}$C, $^{17}$O/$^{16}$O, $^{18}$O/$^{16}$O, $^{33}$S/$^{32}$S and $^{34}$S/$^{32}$S would yield important information, if electronegative elements can be measured by the respective instruments and if the ratios can be determined with sufficient precision.

\subsubsection{Interplanetary dust composition and diversity}
Interplanetary Dust Particles (IDPs) collected in the stratosphere and subsequently analyzed at laboratories on Earth sample a mix of particles from dust-producing bodies. Dust from Jupiter Family Comets is likely dominating~\citep{nesvorny:2010} while dust from the asteroid belt~\citep{rietmeijer:1996} and even the Kuiper belt have also been observed~\citep{keller:2022}. However, stratospheric IDPs are not an unbiased sample of the dust population as their survival during atmospheric entry depends on entry speed and angle~\citep{love:1991}. Measurements of the elemental compositions of IDPs in space would not suffer from this bias and therefore give an average composition of the zodiacal cloud. Further, compositional mapping over different orbital distances could potentially detect differences between Jupiter Family Comets and Kuiper Belt Objects. The major element composition (e.g., Mg/Si, Ca/Si, Fe/Si) of IDP particles is variable and it is unclear if the heterogeneity occurs within or between parent bodies since the origin of individual IDPs collected in the stratosphere is unknown~\citep{bradley:2014}. Compositional mapping of IDPs in the solar system and targeted analyses of individual dust-producing bodies would answer that question. Interstellar Probe may cross cometary dust streams on its path towards interstellar space. Modelling of these streams~\citep{soja:2014} and suitable instrumentation to determine dust compositions and to constrain the dust dynamical properties may provide a statistically relevant dataset for a number of comets and for the sporadic dust background, with increasing distance to the Sun. In particular supporting missions in the solar system may \--- with the current and near-future generation of instrumentation \--- be able to analyze IDPs from different well-known sources~\citep[e.g.,][]{krueger:2020,sterken:2023a, sterken:2023b}. \\
\\
The~\citet{poppe:2016} model predicts that Jupiter-family comet grains dominate the interplanetary dust grain mass flux inside approximately 10 AU, Oort-Cloud cometary grains may dominate between 10 and 25 AU, and Edgeworth-Kuiper Belt grains are dominant outside 25 AU. Mapping the composition of dust over those regions with an interstellar probe, and if possible measuring oxygen isotopes (and other isotopic data) would be very valuable for gaining knowledge on the history of the solar system. 

\subsection{Inner and outer source of Pickup Ions}
Pickup ions (PUIs) were originally assumed to be formerly neutral (mainly H and He) particles of interstellar origin that were ionized in the heliosphere and picked up by the solar wind, where they accelerate to higher energies, presenting a cut-off at roughly twice the solar wind bulk speed. These interstellar PUIs enter the heliosphere in the same manner as ISD does (see Section~\ref{sec:ss_in_lic} and \citealp{kallenbach+2000} and references therein). For a review of in-situ detections of PUIs, see \citet{zirnstein+2022}. \\
\\
Measurements from the Solar Wind Ion Composition Spectrometer (SWICS) on Ulysses \citep{geiss+1995} discovered other species of PUIs (in particular C+, and O+, N+, etc.) from an ``inner source'' near the Sun that had previously been hypothesized by \citet{banks1971}. Two competing mechanisms of generation of these inner source PUIs were proposed: (1) solar wind particles are first embedded in and later released from dust grains close to the Sun \citep{gloeckler+2000}, and (2) energetic neutral atoms (ENAs) are created in the heliosheath and propagate close to the Sun, where they are ionised and picked up by the solar wind \citep{gruntman+2004}. \citet{schwadron+2007} showed that the second mechanism is dominant for inner source PUIs. \citet{szalay+2021} confirmed from measurements by Parker Solar Probe (PSP) that submicron-sized dust grains do not have sufficient cross-sections to produce all inner source PUIs; however, because nanograins can become trapped close to the Sun, they may account for inner source PUIs via mechanism (1).\\
\\
Neither interstellar nor inner source PUIs can explain the presence of easily ionized atoms among the measured composition of anomalous cosmic rays (ACRs)in the outer solar system ( \citealp{cummings+2002}, \citealp{schwadron+2007}). Therefore, an additional ``outer source'' of PUIs was proposed: sputtered atoms from dust grains originating in the Edgewort-Kuiper belt (EKB) are ionised and picked up by the solar wind \citep{schwadron+2002}. \\
\\
The dynamics of nanodust is expected to be similar to that of PUIs but the contribution of nanodust (heavy charged particles that may be multiply charged) to the physics and boundaries of our solar bubble has not currently been quantified. To date, we only have limited knowledge about the PUI distribution in interplanetary space from the New Horizons mission, from older missions/instruments like Ulysses/SWICS, and (in the heliosheath) from indirect measurements of PUIs by Cassini and IBEX that measure remotely sensed Energetic Neutral Atoms (ENAs) from 10 AU and 1 AU, respectively. Measurements from the Solar Wind Around Pluto by the \citep[SWAP;][]{McComas_ea:2008} instrument on the New Horizons mission showed that the interstellar PUIs are heated in the frame of the solar wind, before reaching the termination shock \citep{McComas_ea:2021}. Notably, once the Voyager missions crossed the termination shock \citep{Decker_ea:2005,decker_ea:2008,stone_ea:2005,stone_ea:2008} they identified that the majority of the shocked solar wind energy density went into heating the PUIs, whereas $>$15\% was transferred to energetic ions, showing an unexpected charged particle spectrum inside the heliosheath \citep[e.g.]{dialynas_ea:2019,dialynas_ea:2020}. Only $\sim$20\% of the shocked solar wind energy density went into heating the downstream thermal plasma \citep{richardson_ea:2008}. Consequently, PUIs are expected to play a substantial role in the pressure balance between the heliosheath and the Very Local Interstellar Medium (VLISM), but the Voyager missions could not measure PUIs. The analysis of a unique combination of all available \textit{in situ} ion and remotely sensed ENA measurements \citep{dialynas_ea:2020} over $\sim$10 eV to $\sim$344 MeV energies, showed that the heliosheath is a high plasma-$\beta$ region ($\beta$ is here the particle over the magnetic field pressure), where PUIs (primarily) and suprathermal particles (secondarily) dominate the pressure \citep[see also review article by][]{dialynas_ea:2022}. Understanding both the nanodust and PUI populations through direct in situ measurements from a future ISP mission will be instrumental for understanding the heliosphere's interactions with the Very Local Interstellar Medium (VLISM).

\section{An interdisciplinary science case and its importance for a wider field}
\label{sec:sciencecase}
Here we summarize the most pressing science questions covering the fields of heliospheric (H) and dust science (D), and questions related to the heliosphere-dust interaction (HD). Addressing in depth this broad spectrum of questions is also important for the astrospheric community and for understanding our local interstellar neighborhood. In the following, we divided the questions according to the dust size, so that they can be linked more easily to the type of measurements and instrumentation that are needed. Apart from ISD, interplanetary (nano)dust may also play a role in these questions. \\
\\
{\bf Micron-sized ISD}: What is the gas-to-dust mass ratio in the ISM, and hence, what is the biggest size of ISD residing in the Interstellar Medium (ISM)?$^{(D)}$ Do large grains detected at Earth as (interstellar) meteors exist in the ISM?$^{(D)}$ Is any of the dust coming from a direction other than the heliosphere nose and what does it imply for our current interstellar environment near the interface between LIC and G-cloud?$^{(D)}$ What is the composition and morphology of micron-sized ISD (porous, aggregate, compact?) and what implications are there for the formation of the dust and processes in the VLISM?$^{(HD)}$ What are the characteristics of Oort cloud dust, and what will the Kuiper belt dust reveal about its sources?$^{(D)}$\\
\\
{\bf Submicron-sized ISD}: How do ISD dynamics depend on the heliosphere, and specifically how does the heliosheath filter out these particles?$^{(HD)}$ What is the time-variable size and structure of the heliosphere (using dust measurements as additional boundary conditions for the heliosphere models)?$^{(H)}$ From which distance to the Sun can we measure carbonaceous ISD, and why has there been little evidence in detections so far? \\
\\
{\bf Nanodust ISD}: How much nanodust is filtered (time-dependent or permanently) at the heliopause and heliosheath?$^{(HD)}$ What role does the nanodust inside and outside of the heliopause/heliosheath play in heliospheric physics?$^{(HD)}$ Does nanodust pile-up near the heliopause?$^{(HD)}$ Where does ‘outgoing’ (interplanetary) nanodust from the solar system and the ISD reside in the heliosphere; i.e. will they flow to the heliosphere flanks?$^{(HD)}$ Can it affect the heliosphere size and structure throughout the solar cycle?$^{(HD)}$ What are carbon nanodust species made of and will we measure Polycyclic Aromatic Hydrocarbon (PAH) clusters outside of the heliopause?$^{(D)}$\\
\\
{\bf All dust sizes}: How much charging does ISD acquire in different regions of the heliosphere, in particular in the heliosheath, and how does this charging depend on dust size, composition and local environment properties?$^{(HD)}$ Does dust – and what sizes of the dust – play a role in the pressure balance of the heliosphere?$^{(HD)}$ How does dust affect the production of pickup ions, and how does it depend on the solar cycle? Do ISD or interplanetary dust particles (IDP) contribute to mass-loading of the solar wind?$^{(HD)}$ What are the different dust populations in the ISM, and what are their compositions, particle morphologies, and bulk densities?$^{(D)}$ How do they compare with astronomical measurements and cosmic abundances?$^{(D)}$ How much do they affect the plasma / heliosphere physics, and at which spatial scales?$^{(HD)}$ What species of carbonaceous ISD exist and for which dust sizes and abundances?$^{(D)}$ How much of the ISD is likely recondensed or pristine stardust?$^{(D)}$ How accurately does our current knowledge of elemental and isotopic composition, mostly derived from measurements of the solar nebula and galactic cosmic rays, reflect that of the galaxy/universe?$^{(D)}$ What is the role of the dust for astrospheres?$^{(HD)}$ What is the role of the dust in the history and habitability of the heliosphere?$^{(HD)}$\\
\\
{\bf Importance}: Probing the heliosphere-dust interaction using modelling and in situ measurements is essential for understanding our own immediate interplanetary and interstellar environment. It is also a test-bed to understand how other astrospheres work, as well as to unravel the history of our own solar system and its interaction with various environments during its journey through the Galaxy. Tracers of this journey can now be found in deep-sea sediments (e.g., from supernovae \citep{Miller:2022} or perhaps from passing through denser clouds \citep{opher2022a}. Dust from the VLISM is of particular astrophysical interest in light of recent near-Earth supernovae from which debris is still falling on Earth today \citep{Koll:2019} and must arrive in the form of dust \citep{Miller:2022}. Studying this dust is also important for galaxy evolution and physics of the ISM (see Section~\ref{sec:composition}).

\section{Assessment of infrastructure, research strategy to answer these science questions, and technological development needs}
\subsection{Dust measurements on an Interstellar Probe}
First and foremost, a mission into interstellar space like the Interstellar Probe \citep{mcnutt:2022, brandt:2022} with a dedicated dust detection suite on board would be \underline{optimal} for compelling ISD and heliosphere research. Such an Interstellar Probe (ISP) would \-- for the first time \-- be able to measure the smallest ISD particles beyond the heliopause that are blocked from entering the solar system. With such measurements, ISP would be entering unexplored scientific territory. Also, these dust particles of a few to tens of nanometers are \underline{orders of magnitude} more numerous than the particles Ulysses could measure (see Figure~\ref{fig:ISD_size_distr}). In addition, ISP could detect whether there is really a pile-up of particles near the heliopause. For the first time, we would be able to measure how and until what size the particles follow the flow of the VLISM, which sizes can cross the heliopause (heliopause permeability), and how far some particles can travel through the heliosheath. Such measurements in combination with measurements of the local magnetic field, plasma properties, pickup ions, and the surface charge for dust particles larger than a few hundred nanometers, will help tremendously in understanding the heliosphere-dust interaction and the potential role of dust in heliosphere physics. Also, ISPs move fast (ca. 7-8 AU per year outward) into the stream of ISD (coming at 5.5 AU per year inwards). The high speed results in higher fluxes (cf., detection rates) and enhanced detector sensitivity for the dust impacts, making the detection of tiny particles easier as well as allowing particles to be fully ionized for all compositional elements. Last but not least, ISP will fly throughout approximately 16 years, more than a solar cycle, while passing through interplanetary space, the termination shock, the heliosheath, up to the heliopause and beyond, making it an optimal mission for studying the heliosphere-dust coupling and using this knowledge for other astrospheres. Beyond the heliopause, the tiny dust with gyroradii of only a few to 100 AU (for dust radii $<$~0.1~$\mu$m, see also Table~\ref{tab:charges}), will help study the interstellar environment (magnetic field, plasma) and may detect local enhancements of smaller as well as bigger ISD. The strength of the mission lies in flying through all of these diverse regions with simultaneous magnetic field, dust, plasma and pickup ion measurements. No mission so far has flown a dedicated dust dynamics and composition suite into the heliosheath and the vast space beyond.

\subsection{Continuous observations and observations from different vantage points in space}
The optimal way to disentangle the spatially and temporally variable dust dynamics in the heliosphere is by ensuring long-term monitoring of the dust flux ($>$~22 years) and by combining measurements from different vantage points in space. Hence, the science yield of an ISP mission would be greatly enhanced by simultaneous measurements inside the solar system by another mission, with a dust suite tailored to measuring dust dynamics (and composition) over an extended period of time. One example of such an observing capability in the ecliptic plane is a long-term dust suite on the Lunar Gateway~\citep{wozniakiewicz:2021, sterken:2023a}, with the continuation of complementary dust measurements by IMAP~\citep{mccomas2018}. Examples of such missions out of the ecliptic could be the DOLPHIN(+) mission concept that was proposed to ESA 2022 \citep{sterken:2023b}, the SunCHASER mission concept that includes a dust detection suite in its baseline~\citep{posner:2021}, or a mission with a Ulysses-type orbit that is out of the ecliptic and perpendicular to the ISD stream. Missions with inclined orbits can in addition investigate the IDP-heliosphere interactions, and the solar-cycle dependent vertical structure of the zodiacal dust cloud. Such a dust suite could contain a Large Area Mass Spectrometer \citep{sternovsky:2007, srama:2007} (or a combination with an impact ionization detector), equipped with one or several charge grids / a trajectory sensor, eventually augmented by a large-area polyvinylidene fluoride (PVDF) detector. An in depth overview and discussion of possible ISP instrumentation is given in Section ~\ref{sec:instrumentation}, while an overview table of the main goals of the above mentioned supporting missions is given in Table~\ref{tab:missions_comp}. 

\subsection{Synergies between heliosphere and dust measurements, inclusion of ‘serendipity instruments’, and modelling}
\begin{figure}
	\includegraphics[width=\columnwidth]{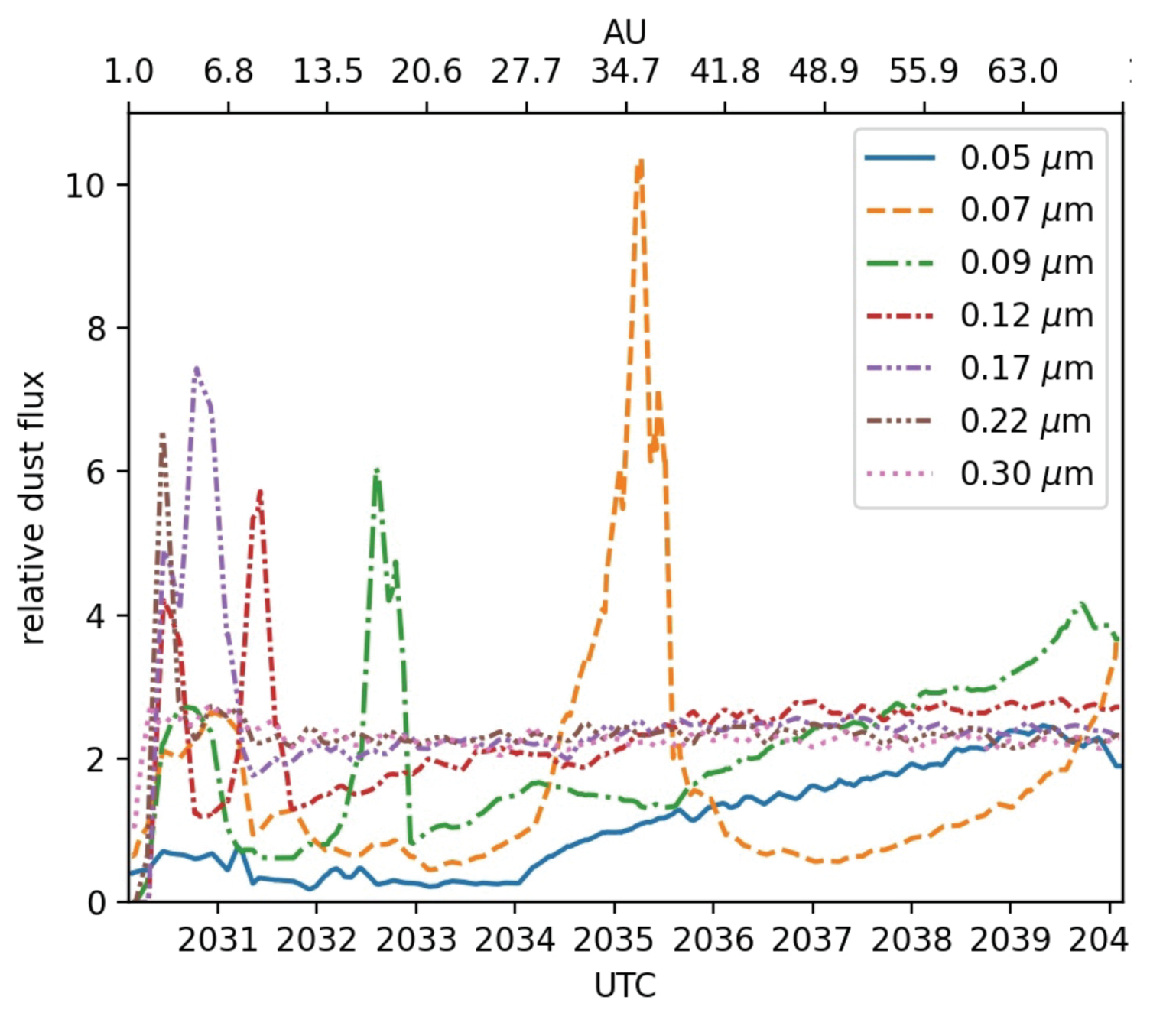}
    \caption{An example of how computer simulations of relative dust fluxes can teach us about the filtering at the heliosheath, when compared to spacecraft data for the respective dust sizes, from \citep{hunziker:2023isp}.}
    \label{fig:rel_flux}
\end{figure}

Simultaneous measurements with complementary instruments, i.e., for plasma and magnetic field properties and pickup ion detections, together with dust fluxes, velocities, directions and – if possible – dust surface charge will yield particularly strong synergies between dust and heliospheric science. The inclusion of `serendipity dust instruments’ that collect information on dust impacts, but were not originally designed for this work, will enlarge the pool of data to be used from different vantage points in space. Plasma wave instruments on various satellites, which pick up a sharp signal when a dust particle impacts the spacecraft, are very good examples of this. The Wind mission yielded a yearly recurring ISD signature in the more than 25 years of plasma wave dust data, including a solar cycle variability \citep{malaspina:2014, malaspina:2016, hervig:2022}. Also Voyager has detected a few impacts \citep{gurnett:1983}. A challenge is that the operations and observations were not tailored to dust impacts; hence, retrieving the dust flux and direction is a challenging task. Also, information such as impact velocity, particle mass or particle charge is missing. Therefore, it is difficult in the solar system to distinguish between interplanetary dust particles (IDP) and interstellar dust (ISD) impacts with these types of instruments. \\
\\
A long-term dust monitoring mission, with sufficiently large detector surfaces, dust trajectory, surface charge, and velocity sensing capabilities (and composition) would be a tremendous leap forward and a significant increment to this pool of data. In any case, Wind has fuel for another 50 years \citep{darling:2019}, IMAP (with dust compositional analyser, without grid) could keep monitoring the compositions and fluxes of incoming ISD on a statistical basis, and the Gateway may be a good platform for long-term monitoring during the flight time of an Interstellar Probe. 
When such a data set (multiple places and long-term) is combined with state-of-the-art computer modelling of the heliosphere-dust flow, then the particle properties (e.g., size distribution in the LIC) and the dynamical structure of the heliosphere can be retrieved by fitting a model of the heliosphere, including time-variable heliosheath, to the pool of data. Figure~\ref{fig:rel_flux} illustrates that even a simple model with only dust filtering in the solar system can already yield valuable information about filtering at the heliosheath if sufficient data are available. The model used is the IMEX model \citep{strub:2019} and the predictions shown are for an interstellar probe. The ISD waves ‘rolling’ in can be seen as sharp increases in relative flux at different times for different particle sizes. An additional filtering at the heliosphere boundaries would alter this pattern. These fluxes are predicted along an ISP trajectory with launch date in 2030, during the focusing phase of the solar cycle. Dust observations along the path of ISP at high impact velocities may be able to shed light on heliospheric filtering, through monitoring whether such patterns are present inside of the heliosphere, in addition to the direct measurements in the heliosheath. Similar investigations can be undertaken in the solar system.

\subsection{Ground-based facilities}
An on-going calibration effort of different dust detectors with a dust accelerator is crucial for success. Since ISP moves very fast, calibrations with a dust accelerator are needed with high velocities and for dust particle analogs with different properties (e.g., lower bulk density dust analogs are important for measurements of $\mu$m-sized ISD \citep{hunziker:2022}). New dust analogs need to be further developed, measurements with plasma wave instruments need to be further understood \cite[e.g.][]{shen:2021} , and high-level computing facilities are needed for the modelling. The dust accelerator at LASP, Univ. Colorado Boulder and at IRS, Univ. Stuttgart, are indispensable tools for any space mission with a dust detector on board. Developments are underway at the University of Stuttgart for a linear staged accelerator (faster velocities), and at ETH Zürich and FU Berlin for the next generation of dust analogs. High-precision and high-power ($>$ 100 kW pulse) ground-based radars are needed for interstellar meteor research \citep{hajdukova:2020}.\\
\\
The technological risk for these types of missions, instruments and ground-based facilities is relatively low, since most have been developed already or are based on heritage. 

\section{Available instrumentation and how they address the science objectives and questions}
\label{sec:instrumentation}
This paper does not propose a mission or an in-depth Science Traceability Matrix, but we discuss instrumentation that is available today or in the near future, its strengths and weaknesses with respect to different types of measurements, and how it can contribute to the science questions and science goals for an Interstellar Probe and support missions. We focus mostly on science questions related to ISD- (or IDP-) heliosphere interactions. Questions focusing on other fundamental physics aspects of the heliosphere, in part as a result of the Voyager 1, Voyager 2, Cassini, IBEX and New Horizons missions, are reviewed in \cite{dialynas_ea:2023} and particularly in~\citet{brandt:2022,brandt_ea:2023} and~\citet{mcnutt:2022}. \\
\\
The science questions in Section~\ref{sec:sciencecase} can be summarized in the following science objectives (SO) (see also Table~\ref{tab:largescience}):\\
(SO1) Nature of our local insterstellar medium environment \\
(SO2) Origins and processes of dust in the local interstellar medium \\
(SO3) Origins and processes of dust in or nearby the heliosphere \\
(SO4) Heliosphere structure, physics and dynamics. 

\begin{table*}
\begin{center}
\begin{tabular}{|p{4cm}|p{2cm}|p{7cm}|} \hline
Dust size regime        & Science Objectives & Main type of questions      \\ \hline \hline 
$\mu$m-sized dust from the ISM   & SO1, SO2    & ISM exploration, sources, processes, populations 
\\ \hline
Electromagnetically dominated (sub-micron) sizes & SO3, SO4   & Heliosphere-dust dynamics, dust as a tracer, heliosphere-dust interactions and processes, ISM dust dynamics   \\ \hline
Nanodust sizes / macromolecules & SO4 & Influence of dust on heliosphere plasma, dust as a tracer in the heliosphere and the ISM, PAH in the ISM   \\ \hline
\end{tabular}
\caption{Summary of science objectives and types of science questions considered in this publication with strong dust-heliosphere synergies.}\label{tab:largescience}
\end{center}
\end{table*}

\subsection{PVDF detectors}
PVDF detectors employ a permanently polarized polymer film that generates a charge pulse upon particle impact. The penetration of the film causes a depolarization of the material, ensuing a measurable relocation of charge \citep[see, e.g.,][]{simpson:1985}.
The shape and amplitude of this signal depend on the mass and impact speed of the dust particle. PVDF sensors are foil-type detectors, named after the material used as the polymer (polyvinylidene flouride).\\
\\
PVDF detectors have the advantage of being low-cost, low-resource, and fairly simple. They can be used to cover large areas \citep[e.g., $0.54\,$m² onboard IKAROS,][]{hirai:2014} 
and may even be integrated with a spacecraft's thermal insulation \citep[e.g., onboard EQUULEUS,][]{funase:2020}.
A student-project PVDF detector currently flies on the New Horizons mission, providing measurements from beyond 55~AU~\citep{horanyi:2008,bernardoni:2022}. PVDF detectors are particularly useful for the micron-sized part of the dust size distribution. However, they lack the possibility to distinguish impact mass and impact speed, and contain no information about impactor directionality except for the pointing of the instrument with a field of view of 180$^\circ$. Due to their piezo- and pyroelectric properties, PVDF sensors can generate noise events induced by mechanical vibrations or thermal variations \citep{simpson:1985,james:2010}. These can be mitigated to some degree by correlation of events with spacecraft operational activities or by use of shielded reference sensors to adjust the trigger threshold \citep{piquette:2019}.\\
\\
PVDF can contribute to the science questions (Section~\ref{sec:sciencecase}) related to large interstellar dust particles
and (towards) interstellar (micro)meteoroids~\citep{gregg:2023, hajdukova:2019}, provided that the spacecraft is outside of the solar system. For a spacecraft in orbit around the Sun, e.g. at Earth distance, the yearly modulation of the dust fluxes \--- due to the fact that ISD from the LIC comes mostly from one direction \--- provides some information about the ISD flux as well~\citep[e.g.,][]{hervig:2022, malaspina:2014}. The inability to discriminate between ISD and IDP with PVDF makes it less useful for studies of ISD inside the solar system. Because PVDF is suitable for micron-sized ISD detections, owing large surface areas, it can contribute to a certain extent to the science questions concerning the gas-to-dust mass ratio in the ISM, finding `big' dust grain populations, and to support astronomical observations of interstellar dust in the micron-size regime and above. 

\subsection{Impact ionisation detector}
When a dust particle impacts the target of an impact ionisation detector (IID) at hypervelocity speeds, particle as well as target material are vaporized and partially or fully ionized (depending on speed). The charge of the generated ions and electrons is measured, and the impact speed and mass of the dust particle are estimated through calibrated signal rise times~\citep[e.g.,][]{grun:1992} and charge signal amplitudes~\citep[see, e.g.,][]{friichtenicht:1963, grun:1992} respectively.\\
\\
Impact ionisation detectors achieve a high degree of reliability and sensitivity, with the ability to detect impactors of only tens of nanometers in size (or below for fast speeds, see also Section~\ref{sec:measuring_nanodust}). Considerable surface areas in the order of~$0.1\,$m² can be accomplished (e.g., onboard Ulysses) despite relatively simple and lightweight designs.
However, they only obtain limited information about the dynamics of the impactors. The directional constraint comes solely from the aperture design~\citep[consider e.g., the FOV half angle of the Cosmic Dust Analyzer (CDA\footnote{Note that CDA was a combined TOF-MS and IID with two charge-sensing grids.}) onboard Cassini of $\sim45^\circ$,][]{srama:2004}. Impact velocity estimates based on the charge signal shape involve large uncertainties in the order of a factor of 1.6--2~\citep{goller:1989}, which may be even higher for fluffy particles \citep{hunziker:2022}. Since the mass of the impactor is derived from the relation Q $\sim$ m\, v$^\alpha$ (with Q the measured charge after impact, m the impactor mass and v the impact velocity), the particle mass uncertainties typically are in the order of a factor of 10~\citep{goller:1989}. This is why reliable velocity information is of crucial importance for impact ionization instruments, in order to distinguish interstellar from interplanetary dust in the solar system, and for estimating the mass-frequency distribution of the interstellar dust in the ISM. Large statistics may nevertheless yield useful information about the dynamics of the dust \citep[e.g.,][]{strub:2015,sterken:2015}. Adding a (segmented) charge-sensing grid or trajectory sensor, however, greatly enhances the science return (see Section~\ref{sec:dusttelescope}). \\
\\
Due to their sensitivity, impact ionisation detectors have been instrumental in the exploration of the smallest meteoroids, such as $\beta$-meteoroids~\citep{wehry:2004}, nanodust (Section~\ref{sec:measuring_nanodust}), and submicron ISD, particularly with regard to their abundance, but also their dynamics~\citep{landgraf:2003, strub:2015, sterken:2015}.
Since these detectors can detect dust from the few-nanometers to micrometer size range, they also are powerful for a larger number of science questions related to the ISD size distribution inside and outside of the heliosphere, the modulation of the ISD dynamics by the heliosphere, the gas-to-dust mass ratio, etc. comparable to the work done with the Ulysses mission dust data, if there is a sufficiently large surface area. The fact that the instrument is sensitive to dust impacts of a few nanometers at relative speeds typical for ISP (ca.\ 55~km/s,~\citealp{hunziker:2023isp}) makes it useful for nanodust studies as well (see also Section~\ref{sec:measuring_nanodust}). Therefore, such instruments can contribute to questions of the dust distributions in and around the heliosphere, including a possible pile-up of ISD outside the heliopause, and consequences for the physics of the heliospheric boundary regions. 
These dust measurements throughout the solar system may be used as an extra boundary condition for the heliospheric model, but inside of the heliosphere, it is challenging to discriminate ISD from IDP. \\
\\
Adding a segmented grid (see Sections~\ref{sec:grid} and~\ref{sec:dusttelescope}) for better velocity determination would greatly augment the science return both from the point of view of distinguishing populations as well as for having more precise estimates of the particle masses. Since the instrument is more sensitive than PVDF, it would also augment our current knowledge on dust in the Kuiper belt region (in particular if combined with a grid), after the first crude dust measurements in the region were taken by the Voyager mission using plasma wave antennas~\citep[][pers. comm.]{jaynes:2023}, and by the New Horizons mission using a PVDF detector~\citep{bernardoni:2022}. Although this type of fairly simple and well established instruments can contribute to many science questions concerning populations, dynamics and in particular dust-heliosphere interaction and physics, it still lacks compositional information for many of the origins, processes and populations related science questions. Compositional information can also help discriminate ISD from IDP. A combination with a time of flight mass spectrometer (Section~\ref{sec:tof-ms}) and/or (segmented) grid can be flown (e.g. Cassini CDA).

\subsection{Time of flight mass spectrometer (TOF-MS)}
\label{sec:tof-ms}
Dust particles impacting at hypervelocity speeds on a time-of-flight (ToF) mass spectrometer ionize by the impact ionization process. The plasma's ions are then accelerated by an electric field (e.g., 1000~V for Cassini CDA) that separates them
according to their charge-to-mass ratios. The recording of their travel times from the impact target to an ion detector then yields the abundance of species with different charge-to-mass ratio within the impact plasma, from which the information about the impactor composition can be derived.\\
\\
The key benefits of the TOF-MS are: (1) its ability to analyse the grain composition, and (2)
that the recording of a mass spectrum functions as unequivocal proof of a true particle impact
(as opposed to a noise event).
Higher mass resolutions can be obtained with more sophisticated mass analyzer concepts (`ion-optics') than for the linear TOF-MS:
\begin{itemize}[leftmargin=*, labelindent=1em]
    \item linear; $m/dm \approx  30$; e.g. CDA, \citep{srama:2004}
    \item reflectron; $m/dm \approx 100\!-\!300$; e.g. DDA, SUDA \citep{kruger:2019,kempf:2018}
    \item orbitrap; $m/dm > 10,000$; e.g. onboard SLAVIA \citep{zymak:2023}
    yet to be tested in space
\end{itemize}

\noindent One limitation of the TOF-MS is that only either the plasma's cations or the anions can be fed into the mass analyser, depending on the polarity of the ion optics. So far only cation mass analysers have been used in space missions, as cations are readily formed by most elements and molecules \citep{srama:2009}. Certain organic molecules, however, form anions rather than cations during impact ionisation \citep{hillier:2014,hillier:2018}, suggesting the use of (switchable) dual polarity ion optics in future instruments \citep[as first employed in the upcoming SUDA instrument,][]{napoleoni:2023}. Two arguments especially support an anion analysis of impact plasma of a dust particle: (1) oxygen can be measured with a much higher sensitivity (up to factor 10$^5$) which would allow the determination of isotopic ratios. This is important for the sensitive detection of water ice, hydroxides, silicates and oxides. However, its yield for cations is strongly impact speed dependent. (2) A negative anion mode would also allow a sensitive study of Halogens, Carbon and minerals like S, P, SO$_4$ and PO$_4$. This complements the sensitive detection of metals in the cation mode. In summary, using the combination of cation and anion modes in TOF-MS impacts with speeds above 30 km/s allows the sensitive detection of all elemental ions between 1 and 200 amu.\\
\\
Isotopes help to identify elemental species, but are not trivial to measure. Measurements of isotopes at mass M require both a mass resolution higher than M and a high dynamic range in order to quantify small peaks in the vicinity of larger peaks. When the dynamic range reaches 1000 or better, the identification of isotopically anomalous interstellar dust grains of circumstellar origin is achievable, provided extensive calibration data is available. The current and former generations of impact ionization TOF-MS were not optimized for simultaneous high-dynamic range and high mass resolution. Future instruments will employ improved electronics in order to extend the dynamic range.\\
\\
Impact velocities also play a major role in TOF impact ionization spectrum analysis. At lower velocities, not all of the impactor constituents may become ionised. Table~\ref{tab:velocities_comp} shows the minimum impact velocities that are needed for the detection of the ion species (for the positive ion mode). Considering the flow speed of ISD of about $25\,$km/s, such velocities are met in most conceivable cases. In particular for ISP, moving into the nose direction of the heliosphere, all particles are expected to be fully ionized at relative speeds of ca. 55 km/s. However, for spacecraft in the solar system moving in the down-wind direction (along their heliocentric orbit or on a down-wind escape trajectory), the relative velocities of ISD may be insufficient for complete ionisation of certain species.\\
\begin{table}
\begin{center}
\begin{tabular}{cc} 
\hline
Species & V$_{\rm min}$ (km/s)        \\ \hline \hline
H       & 8 \--- 10        \\ 
C       & 10 \--- 12       \\
O       & 14 \--- 16       \\
Na      & 2 \--- 5         \\
K       & 2 \--- 5         \\
Mg      & 5 \--- 10        \\
Al      & 5 \--- 10        \\
Si      & 10 \--- 15       \\
Ca      & 5 \--- 10        \\
Fe      & 10 \--- 15       \\
Rh      & 8                \\
S or O$_2$ & 15 \--- 20    \\ \hline
\end{tabular}
\caption{Minimum impact velocities for measuring the composition of the tabulated species with impact ionization time-of-flight mass spectrometry (in positive ion mode). The Mg-peak can also appear even at lower speeds around 3~km/s but is then very small. The Si-peak can also appear at 7 to 10 km/s, albeit also very small. The peaks of S and O$_2$ are difficult to distinguish.} \label{tab:velocities_comp}
\end{center}
\end{table}
\\
Compositional information is crucial for many of the science questions related to the origins and processes of ISD in the VLISM, dust origins and processes in the solar system (e.g. Kuiper belt, various comets), charging mechanisms, generation of PUI, etc. (see also Section~\ref{sec:composition}), but it can also help to discriminate between different dust populations. Also, for such instruments statistics are vital for the science results and hence, the need for large surface area. Large Area Mass spectrometers have been developed with surface areas of 0.1~m$^{\rm2}$~\citep{sternovsky:2007, srama:2007}. Speed information can be constrained within boundaries from the shapes of the peaks, and from the occurrence of the peaks or from molecular clusters. Particle masses can be constrained from the impact charge together with simulations of the ion optics and calibration data. However, a (segmented) grid or trajectory sensor would be of great added value. 
\begin{figure}
    \centering
    \includegraphics[width=\columnwidth]{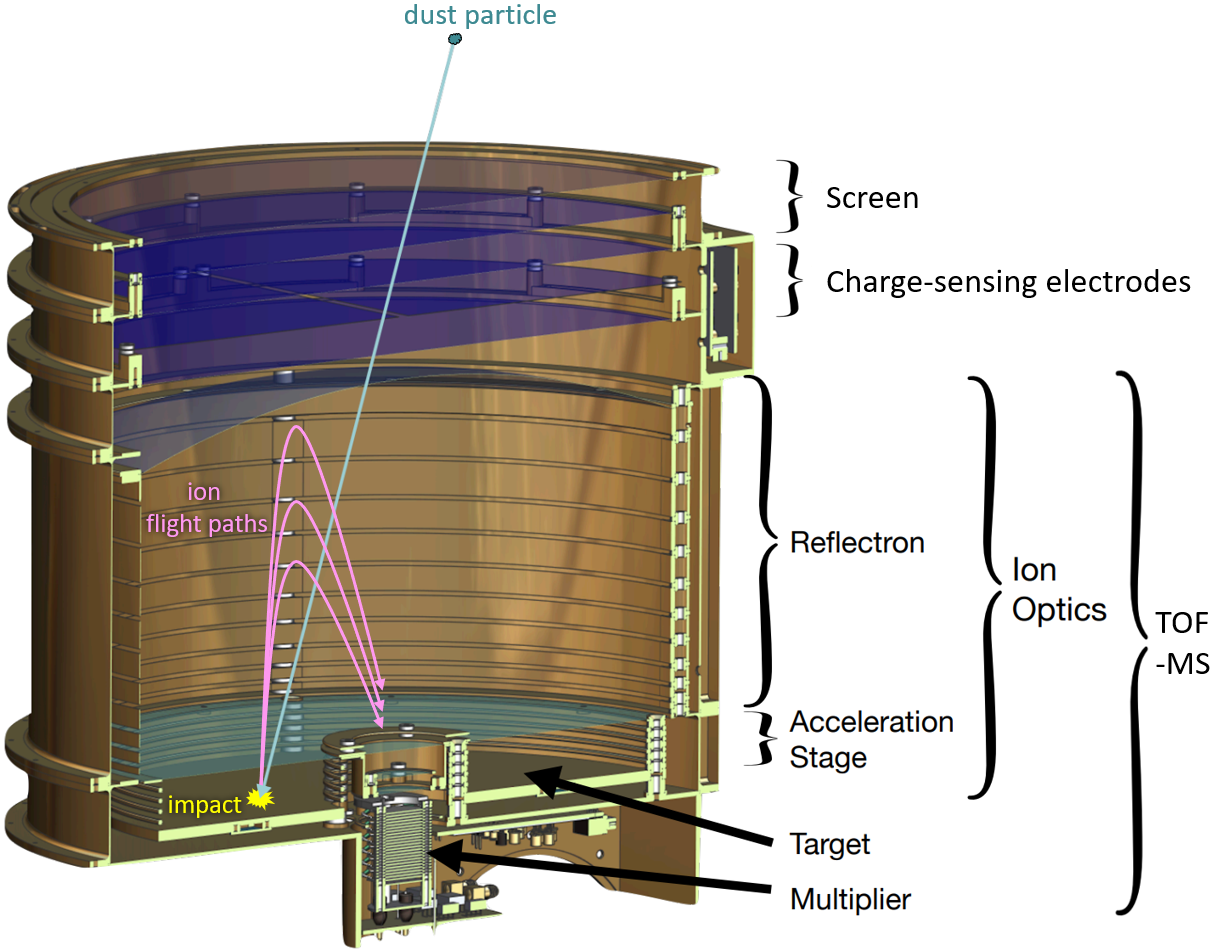}
    \caption{Schematic of the DDA instrument. Image credit: IRS/Univ. Stuttgart}
    \label{fig:DDA_schematic}
\end{figure}
\subsubsection{Plasma Wave Antennas}
When a dust particle impacts the body of a spacecraft at hypervelocity speeds, it vaporises both itself and a fraction of the spacecraft surface. The recollected electrons and the induced charges of the escaping electrons and ions measured by the plasma wave antennas produce a distinct amplitude signal. Whether the impact speed and mass of the dust particle can be reconstructed from that signal is currently a topic of investigation \citep[see, e.g.,][and references therein]{shen:2023}.\\
\\
The advantage of the plasma wave antennas is the large surface area (basically, the spacecraft body), and the science-at-no-extra-cost if a plasma wave antenna is on board. However, it comes with the caveat that only limited directionality information can be derived (e.g.,~\citep{malaspina:2014, pusack:2021}). Also, the signal is a function of mass, impact velocity, distance to the antenna (depending on configuration), and spacecraft surface material. As a consequence, it is not yet possible to uniquely determine the mass, velocity and/or distance to the antenna of the impact. A plasma wave antenna can detect many dust impacts per time period compared to other dust instruments. This information can be heavily compressed into a low-bandwidth data product via on-board dust detection algorithms. \\
\\
Outside of the heliosphere, the plasma wave antennas can be especially useful because large counting statistics (due to the large surface area of the spacecraft body) may yield important information on the distribution and populations of ISD in the ISM.  Large counting statistics are especially useful for detecting larger, more sporadic dust grains. However, just like PVDF detectors, only limited information on mass and velocity can be derived. Inside of the heliosphere, at e.g.\ Earth orbit, the plasma wave antennas can infer information about the ISD variability with time through the modulation of the flux throughout the year and throughout the solar cycle~\citep{hervig:2022, malaspina:2014}. Plasma wave and PVDF results could be compared with each other.

\subsection{Charge-sensing grid}
\label{sec:grid}
Charge-sensing grids are grid electrodes that sense charged dust particles passing through, via the charge they induce in the electrode (e.g., see upper elements in Fig.~\ref{fig:DDA_schematic}). In addition to measuring the particle charge, dynamical information such as entrance angles and speed may be estimated from the signal shape of the induced charge. \\
\\
Different configurations of charge-sensing entrance grids have been proposed. The Cassini Cosmic Dust Analyzer (CDA) used a serial electrode design with two canted grids that could yield speeds as well as incident angles~\citep{auer:2002}. However, the large capacity of the two grids restricted this design to relatively large grains with charges $>1\,$fC. A design employing segmented, lower-capacity grids has been proposed by~\citet{li:2014,li:2015,li:2017}. Such a segmented design is awaiting in-flight demonstration on-board of the Destiny+ mission\footnote{Demonstration and Experiment of Space Technology for INterplanetary voYage Phaethon fLyby and dUst Science, to be launched in 2024}~\citep{arai:2018} as part of the Destiny+ Dust Analyzer (DDA)~\citep{simolka:2022} with an anticipated detection threshold of $0.2\,$fC. This corresponds to a dust particle with radius of 0.35 $\mu$m (in the solar system). Segmented charge sensing grids are a good compromise between increased science output and instrument complexity. Since they are non-destructive, they are especially suited to be combined with destructive detector stages, so that even a single-plane charge-sensing grid can be used for time-of-flight impactor speed measurements (as done in DDA). The DDA system can determine the speed with a ca. 15\% and the mass with approximately a 20\% accuracy. \\
\\
Charge-sensing grids are well suited to study the abundance and dynamics of bigger ($>1\,$µm diameter) particles in the solar system. These particles are on the larger end of the dust particle size range that is still affected by electromagnetic forces in the solar system. Measuring or constraining their speed would increase the accuracy of the mass determination (Q $\sim$ m $\cdot$ v$^\alpha$), and constraining the speed and velocity vector to a certain extent would allow for a better discrimination between the sources of the dust particles (in particular ISD vs. IDP) in the solar system. Measuring their surface charge (with the grid), their mass (through an IID or TOF-MS) and plasma parameters (plasma instrument) may even yield constraints on their bulk densities. In the heliosheath, the dust is expected to reach higher equilibrium potentials of ca.~+6~to~+12V~\citep{kimura:1998} or +8~V~\citep{slavin:2012}, as opposed to ca.~+5V for the solar system (see Table~\ref{tab:charges}). These are equivalent to dust particle radii of ca.~0.3,~0.15 and~0.2~$\mu$m, respectively, which is well within the range of electromagnetically affected dust that reacts dynamically to the solar cycle through the time-variable heliospheric magnetic fields. In interstellar space, the dust particles are expected to have lower charges, corresponding to ca. +0.5V~\citep{gruen:1996} equilibrium surface potential (equivalent dust radii ca.~3.5~$\mu$m). However, it can be expected that micron-sized ISD may be porous~\citep{westphal:2014, sterken:2015}. Assuming a compactness factor of ca. 1/3$^{\rm rd}$, they can be ca. three to four times more charged~\citep{ma:2013}, i.e. they would be detectable by the grid from a radius of ca.~2~$\mu$m (detection threshold 0.2 fC). Adding a charge grid to the design of an IID or TOF-MS (Section~\ref{sec:dusttelescope}) may thus yield important information on the dynamics of the mid-sized (sub-micron) particles in the heliosphere and, in particular, in the heliosheath, as well as help constrain the direction of motion of very large grains or micrometeoroids that may exist in the ISM, in addition to possibly constraining their bulk material densities. Science questions inside the heliosphere (in particular the heliosheath) related to the dynamics and mass distribution of submicron ISD would be easier to tackle if an IID or TOF-MS instrument includes a (segmented) grid, with limited add-on complexity. Such instruments provide higher constraints on dust directionality, velocity (hence, ISD-IDP discrimination), better mass constraints (using the velocity constraints) and useful information on the dust surface charge.

\begin{table*}
\begin{center}
\begin{tabular}{|c|c|c|c|c|c|} 
 \hline
Radius ($\mu$m)   & density (g$\,$cm$^{-3}$)    & mass (kg)     & surface potential (V) & charge (fC) & gyroradius (kAU)        \\ \hline \hline
\multicolumn{6}{|c|}{Interplanetary conditions} \\ \hline
0.1  & 2.5  & $1\times10^{-17}$    & +5     & $0.06$  & 0.5 \\
0.2  & 2.5  & $8\times10^{-17}$    & +5     & \cellcolor{yellow!25}$0.1$   & 2   \\
0.5  & 2.5  & $1\times10^{-15}$    & +5     & \cellcolor{yellow!25}$0.3$   & 9   \\
1    & 0.8  & $3\times10^{-15}$    & +20    & $2$     & 4   \\
5    & 0.8  & $4\times10^{-13}$    & +20    & $11$    & 97  \\ \hline\hline
\multicolumn{6}{|c|}{Heliosheath conditions} \\ \hline
0.1  & 2.5  & $1\times10^{-17}$    & +8     & $0.09$  & 0.7 \\
0.2  & 2.5  & $8\times10^{-17}$    & +8     & \cellcolor{yellow!25}$0.2$   & 3   \\
0.5  & 2.5  & $1\times10^{-15}$    & +8     & $0.4$   & 17  \\
1    & 0.8  & $3\times10^{-15}$    & +32    & $4$     & 5   \\
5    & 0.8  & $4\times10^{-13}$    & +32    & $18$    & 150 \\ \hline\hline
\multicolumn{6}{|c|}{Local interstellar conditions} \\ \hline
0.1  & 2.5  & $1\times10^{-17}$    & +0.5   & $0.006$ & 0.1 \\
0.2  & 2.5  & $8\times10^{-17}$    & +0.5   & $0.01$  & 0.5 \\
0.5  & 2.5  & $1\times10^{-15}$    & +0.5   & $0.03$  & 2   \\
1    & 0.8  & $3\times10^{-15}$    & +2     & \cellcolor{yellow!25}$0.2$   & 1   \\
5    & 0.8  & $4\times10^{-13}$    & +2     & $1$     & 27  \\ \hline
\end{tabular}
\caption{Approximate size, mass, charge, surface potential, and gyroradius, adapted from~\citet{gruen:1996}, assuming spherical particles, a magnetic field strength of 1\,nT in the interplanetary medium, 0.1\,nT in the heliosheath, and 0.5\,nT in the LISM, and a relative particle speed of 400\,km/s in the interplanetary medium, 100\,km/s in the heliosheath, and 5\,km/s in the undisturbed LISM. Gyroradii are upper limits for particle motions perpendicular to the magnetic field. Surface potentials for 0.1~$\mu$m do not take into account the small particle effect and may be larger in reality; for micron-sized particles a compactness of $33\%$ was assumed, which increases the surface potential by a factor of about 4 \citep{ma:2013}. The instrument threshold is indicated in yellow.}
\label{tab:charges}
\end{center}
\end{table*}

\subsection{Trajectory sensor}

\begin{figure}
    \centering
    \includegraphics[width=0.9\linewidth]{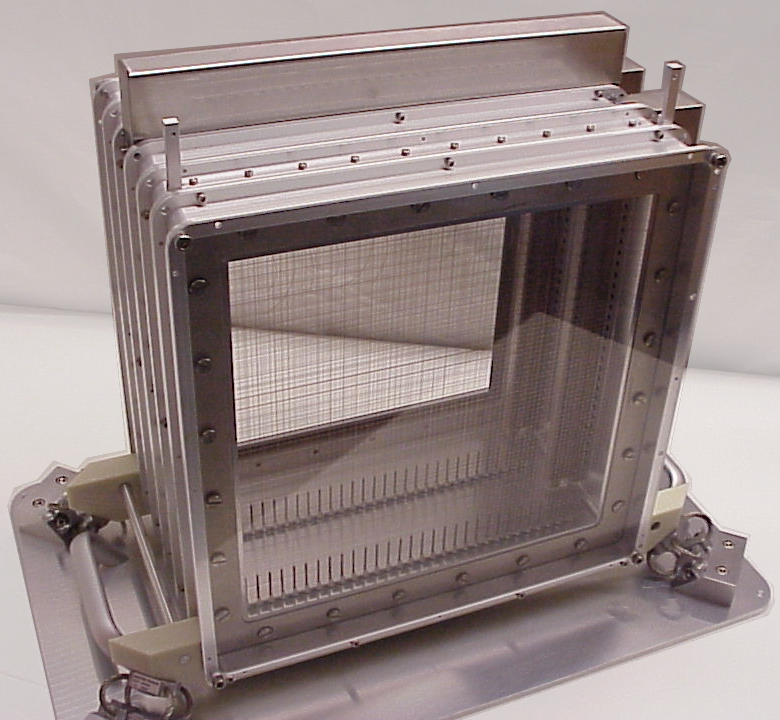}
    \caption{Photo of a trajectory sensor. Image credit: MPI-K/Univ.\ Stuttgart.}
    \label{fig:trajectorysensor}
\end{figure}

The concept of the trajectory sensor involves two planes of position-sensitive charge sensors, from which the flight path of a particle may be accurately reconstructed. These position-sensitive charge sensors can be realized through a set of crisscrossed wire electrodes~\citep{auer:1975,auer:2008} or through a finely segmented grid \citep{li:2014}.\\
\\
The key advantage of a trajectory sensor is its accuracy. For instance, uncertainties of $<1^\circ$ are reached the design of~\citet{auer:2008}. The difficulty of this design lies in its complexity, as one charge-sensitive amplifier (CSA) is required for each wire or grid segment (e.g., 64 CSAs in the design of \citealp{auer:2008}).
So far, such designs (see, e.g., Fig.~\ref{fig:trajectorysensor}) have only been demonstrated in the laboratory \citep{xie:2011}.\\
\\
One of the major motivations for using trajectory sensors, apart from dust surface charge measurements, would be the dynamic differentiation between dust types (e.g., between interstellar and interplanetary dust, or cometary dust streams). Their use together with IID or TOF-MS detectors (see Section~\ref{sec:dusttelescope}) would yield an improved mass determination through an accurate velocity determination (see also the discussion above for the grids).

\subsection{The Dust Telescope}
\label{sec:dusttelescope}
The combination of a non-destructive trajectory sensor with an impact plasma mass spectrometer allows for the simultaneous analysis of dust particles' physical, chemical, and dynamical properties.
This type of instrument as been nicknamed `dust telescope' \citep{grun:2005,srama:2004-2}.
A simplified version of a dust telescope could consist of instruments such as CDA and DDA with their less accurate, charge grid-type dynamics-sensing detector stages. A laboratory model of a true dust telescope (i.e., with high-accuracy trajectory sensor) has been implemented by \citet{horanyi:2019}.

\subsection{Plasma instrument}
The purpose of a plasma instrument (Plasma Subsystem-PLS; see also Table \ref{tab:technicaldata2}) is to measure the low energy (eV to keV) particle distributions throughout the heliosphere with the sensitivity to detect the very cold plasma populations in the VLISM and the dynamic range to measure the solar wind. The physics of the boundaries of our solar bubble, namely the termination shock (TS), the heliopause (HP) and the heliosheath, including the very local interstellar medium (VLISM)~\citep[e.g.][]{dialynas_ea:2022,kleimann_ea:2022} requires the determination of the composition of ions (and electrons) that are "frozen in" to the magnetic field along with an accurate determination of their energy distributions and moments (temperatures, densities, velocities and pressure). Dust in the Solar System is embedded in the solar wind plasma and the relative abundance of dust populations throughout the heliosphere affects the interplay of outflowing solar wind plasma and inflowing interstellar material. Also, measuring plasma parameters along with the dust allows calculating and studying the dust charging process, and constraining the dust bulk densities (morphology) via its surface charges and mass measurements. Direct measurements of the distribution functions of both the interstellar cloud and dust cloud pickup ions up to energies of a few keV/e would be possible with a plasma detector with a geometry factor of $\sim$10$^{-3}$ cm$^{2} \cdot$sr and a signal to noise ratio of $>$10 \citep{mcnutt:2021}.

\subsection{Pickup Ion instrument}
As explained in Section 1.6 and throughout this manuscript, PUIs play a substantial role in the dynamics of our solar bubble, are very important indicators of the plasma processes throughout the heliosphere, i.e. from interplanetary space out to the heliopause \citep[e.g.]{zirnstein+2022,dialynas_ea:2022}, and may be related to dust populations~\citep{schwadron2002+}. Despite the SWAP \citep[]{McComas_ea:2008} and PEPPSI \citep[]{mcnutt_ea:2008} instruments on New Horizons being operational for many years, this spacecraft is not expected to make measurements at distances far beyond the termination shock. Also, its instruments were not designed to measure multiple and heavier species of PUIs (only limited to hydrogen and helium) and there is limited directional information. Furthermore, the limited scientific payload of New Horizons (e.g. it does not carry a magnetometer) indicates that those PUI measurements it obtains cannot be set in context with simultaneous fields or wave measurements. To understand the important physics of our heliosphere through the PUIs and their possible link to dust, a future ISP mission should include a detector with a fairly large geometrical factor ($>$10$^{-3}$cm$^2$ sr), a high dynamic range (10$^{-1}$ to 10$^{4}$ (cm$^2$ sr s keV)$^{-1}$), and a combination of high time and energy resolution (of $\Delta$E/E $\le$10\%) that would resolve light and heavy ions and their charge states within the energy range of $\sim$0.5-78 keV/e \citep[see][]{mcnutt:2021}. 

\subsection{Magnetic field instrument}
The Voyager mission survey through the heliosphere showed that taking accurate magnetic field measurements from interplanetary space all the way to the VLISM is of paramount importance for addressing timeless questions concerning the shape of the global heliosphere, its nature, dynamics and interactions with the VLISM. The relatively low resolution of the MAG experiments on the Voyagers demonstrated the necessity of obtaining magnetic field observations from a future Interstellar Probe mission in the nT range with pT resolution \citep[see][and Table \ref{tab:technicaldata2}]{mcnutt:2021} to address questions concerning the role of plasma turbulence and magnetic reconnection throughout the heliosphere. A high dynamic range (ca.~$\sim$0.01-100~nT) would provide invaluable aid in determining the interaction of small dust grains with particles and fields throughout the heliosphere (e.g., CMEs) and, in particular, in the heliosheath. 

\subsection{Neutral Mass Spectrometer}
The primary science goal of the Neutral Mass Spectrometer (NMS) is to measure the chemical composition of neutral gas along the spacecraft trajectory, employing two measurement techniques: an antechamber and a collection foil. The latter provides a higher sensitivity than the antechamber, but less frequent measurements. The technology readiness level and the longevity of NMS are backed up by e.g. the Neutral Ion Mass Spectrometer NIM \citep{foehn21} on the Jupiter Icy Moons Explorer (launched in 2023, nominal end of mission 2035). NMS may detect dust grains that happen to enter the antechamber or hit the collection foil. However, the collection area is rather small (on the order of cm$^{2}$) compared to dedicated dust detectors, which implies low detection rates. The volatile component of any dust particle entering the antechamber can be measured by the NMS. When nanograins impact the collection foil, both their volatile and refractory species can be analyzed. Additional on-ground calibration at impact speeds representative for the Interstellar Probe will be needed \citep{mcnutt:2021}. Although NMS can be very valuable for the compositional analysis of nanodust and macromolecules in the VLISM, it does not provide impact rates, sizes or dynamical information about the nanodust that would be useful for further exploring the dust-heliosphere physics and the smallest populations of condensed matter in the VLISM.\\
\\
Table~\ref{tab:technicaldata1} and~\ref{tab:technicaldata2} show an overview of the instrumentation discussed, with typical values for measurement ranges, power consumption, instrument mass and volume.

\begin{table*}
\begin{tabular}{|c|c|c|c|c|c|c|}
\hline Dust Parameter & \multicolumn{6}{c|}{ INSTRUMENT } \\ \hline
 & PVDF & IID & TOF-MS & Grid & Traj.-Sens. & Plasma Wave \\ \hline
\hline $\begin{array}{l}\text { Mass, kg} \\
(\mathrm{at}\; 10\,\mathrm{km/s})\end{array}$ & $>10^{-14}$ & $>10^{-17}$ & $\begin{array}{c}10^{-18} \\
\mathrm{to}\; 10^{-15}\end{array}$ & $>5 \times 10^{-15}$ & $>10^{-15}$ & $>10^{-14}$ \\
\hline $\begin{array}{l}\text { Mass, kg} \\
(\mathrm{at}\; 50\,\mathrm{km/s}) \\
\end{array}$ & $>10^{-14}$ & $>10^{-21}$ & $\begin{array}{c}10^{-22}\\
\mathrm{to}\; 10^{-17}\end{array}$ & $>5 \times 10^{-14}$ & $>10^{-15}$ & $>10^{-19}$ \\
\hline Speed, km/s & $1-100$ & $2-100$ & $2-100$ & $2-50$ & $2-50$ & $4-100$ \\
\hline FOV, sr & $2\,\pi$ & $1\,\pi$ & $0.2\,\pi$ & $1.5\,\pi$ & $0.5\,\pi$ & $4\,\pi$ \\
\hline Sens. area, m$^2$ & $\begin{array}{c}\text { variable } \\
0.005-0.1\end{array}$ & $\begin{array}{c}\text { variable } \\
0.05-0.1\end{array}$ & $\begin{array}{c}\text { variable } \\
0.05 \text { - } 0.03\end{array}$ & $\begin{array}{c}\text { variable } \\
0.005-0.1\end{array}$ & $\begin{array}{c}\text { variable } \\
0.01-0.1\end{array}$ & $(0.25-1)$ \\
\hline Observables & $E_\mathrm{kin}$ & $v,\,m$ & $v,\,m$, comp. & $v,\,m,\,q$ & $v,\,m,\,q$, direc. & $E_\mathrm{kin}$ \\
\hline Advantages & $\begin{array}{c}\text { low mass, } \\
\text { low power, } \\
\text { robust }\end{array}$ & $\begin{array}{c}\text { reliable, } \\
\text { sensitive, } \\
\text { nano-grain } \\
\text { detection, } \\
\text { high dynamic } \\
\text { range }\end{array}$ & $\begin{array}{c}\text { composition, } \\
\text { high reliablility, } \\
\text { nano-grain } \\
\text { detection }\end{array}$ & $\begin{array}{c}\text { get } v, m \text { and } q \\
\text { with small } \\
\text { errors }\end{array}$ & $\begin{array}{c}\text { get } \vec{v},\,m,\,q \\
\text { with small } \\
\text { errors }\end{array}$ & $\begin{array}{c}\text { 2in1 } \\
\text { instrument, } \\
\text { large FOV }\end{array}$ \\
\hline Disadvantages & $\begin{array}{c}\text { calibration, } \\
\text { not suited } \\
\text { for inner } \\
\text { solar system }\end{array}$ & errors in $v,\,m$ & $\begin{array}{c}\text { complex } \\
\text { instrument, } \\
\text { HV needed, } \\
\text { limited sens. } \\
\text { area,  } \\
\text { FOV limited,  } \\
\text { errors in }v,\,m \end{array}$ & $\begin{array}{c}\text { sensitive to } \\
\text { plasma, } \\
\text { SNR } \\
\text { for submicron } \\
\text { grains }\end{array}$ & $\begin{array}{c}\text { sensitive to } \\
\text { plasma, } \\
\text { many signal } \\
\text { channels, } \\
\text { power }\end{array}$ & $\begin{array}{c}\text { calibration, } \\
\text { no composition, } \\
\text { no directionality }\end{array}$ \\
\hline Mass, kg & $1-2.5$ & $2-8$ & $4-12$ & $2-5$ & $2-8$ & 37 (RPWS) \\
\hline Power, W & $1-2.5$ & $3-7$ & $5-15$ & $3-6$ & $5-10$ & 16 (RPWS) \\
\hline Data vol., MB/d & $<1 $ & $<1 $ & $1-100 $ & $1-2 $ & $1-100 $ & $2 -1000$ \\
\hline TRL & 9 & 9 & 9 & 9 & $4-9$ & 9 \\
\hline Cost, M€ & 5 & 5 & 10 & 5 & 6 & (12) \\
\hline
\end{tabular}
\caption{Technical data of state-of-the-art dust instrumentation. The mass and velocity range of TOF-MS, grid and trajectory sensors can shift towards larger values depending on gain settings and some instrument adaptations.}
\label{tab:technicaldata1}
\end{table*}

\begin{table*}
\begin{center}
\begin{tabular}{|c|c|c|c|c|} 
 \hline
 & NMS & PLS & MAG & PUI        \\ \hline \hline
Measurement range & 1-1000 amu (molecules) & $<$3 eV/e to 20 keV/e ($\Delta$E/E $\le$ 10\%) & 0.01–100 nT (three components)& $\sim$0.5-78 keV/e \\
Power consumption & 11 W & 10 W & 5.7 W, including two survival heaters & 7 W \\
Mass & 10 kg & 8 kg & 0.6 kg for two fluxgates, 4.2 kg for 10-m boom & 5.5 kg \\
Volume & $40 \times 15 \times 15$ cm$^{-3}$ & -- & -- & -- \\ \hline
\end{tabular}
\caption{NMS, PLS, MAG and PUI specifications according to \citet{mcnutt:2021}.}
\label{tab:technicaldata2}
\end{center}
\end{table*}

\subsection{Discussion on nanodust measurements}
\label{sec:measuring_nanodust}
A fundamental question for in-situ instrumentation is what the lower detection limit in particle size is. Fortunately, the detection method of impact ionization is extremely sensitive to small particles as long as the impact speed exceeds a certain limit. Above impact speeds of approximately 30\,km/s, the particle becomes fully ionized, providing enough ions to be detected with sensitive ion detectors. \\
\\
The measurement of nanodust with sensitive non-TOF detectors (\textit{Galileo}, \textit{Ulysses}, \citealp{gruen:1993b}) and with TOF-MS instruments (\textit{Cassini}, GIOTTO) is well known and published. \citet{utterback:1990} identified particles of only $5\times 10^{-22}$\,kg during the flyby of comet Halley in 1986 with the mass spectrometer PIA/PUMA. The relative flyby speed was 78\,km/s and the smallest signals contained only 75 ions from the generated impact plasma.\\
\\
The instruments onboard \textit{Galileo} and \textit{Ulysses} detected the Jovian dust streams. Models have shown that these particles reach 400\,km/s with typical particle sizes between 10 and 20\,nm \citep{zook:1996}. These detectors used large target areas of up to $0.1\,\mathrm{m}^2$, although their large targets and related large electrode capacitances led to a low sensitivity. Later, \textit{Cassini} characterized the Jovian and Saturnian dust stream particles, measuring the composition of fast and tiny grains: Saturnian stream particles typically have speeds between 100\,km/s and 200\,km/s and are usually smaller than Jovian stream particles \citep{kempf:2005,horanyi:2000,hsu:2011}. Another good example of measuring the composition of individual grains smaller than 50\,nm at moderate impact speeds of approximately 30\,km/s is the \textit{Cassini} CDA proximal orbit campaign with its inner ring plane crossings in 2017 \citep{hsu:2018}. This demonstrated a high sensitivity for simple TOF-MS instruments using impact ionization. \\
\\
Not only dust spectrometers were able to detect nano-sized grains. \citet{carpenter:2007} demonstrated the measurement of nanometer-sized dust impacts with an instrument combining a thin foil and a multichannel plate (MCP) in Earth orbit onboard the ISS.\\
\\
In order to determine the lower mass threshold in dependence of the impact speed, one can use calibration equations by \citep{gruen:1984} of $Q = 6.3\times 10^{-4} \cdot m \cdot v^{5.6}$, or \citet{burchell:1999} with $Q = 0.096 \cdot m \cdot v^{4.01}$. Using the Burchell equation gives a total impact plasma charge of $Q = 6.2\times 10^{-16}$\,C for a 50\,km/s and $10^{-21}$\,kg particle (4.6\,nm radius, silicate). However, at such high speeds, it is a good approximation to assume full ionization of the projectile material. Furthermore, the plasma is dominated by ions from the target material (with or without surface contaminations). If the goal is not the detection of a particle but a careful compositional analysis or mass determination of the nano-meteoroid, one should consider only the dust particle material, and the typical equations giving $Q/m$ for a given size and speed are not applicable. An estimate of the number of dust particle atoms gives, therefore, a more precise quantity of the relevant impact charge. \\
\\
We consider the sensitivity for an impact ionization detector with ion optics that focuses all generated ions towards an ion detector that could be either an MCP or a multiplier. The target is normally a polished metal surface (gold, rhodium, iridium, palladium) onboard dust telescopes like SUDA (\textit{Europa Clipper}) or DDA (DESTINY$^+$). The loss factor from the target to the ion detector is assumed to be 50\%. An MCP and multiplier is sensitive enough to measure even single ions. But such high detector gains are not practical in space due to the high sensitivity to radiation. Therefore, the ion detector should operate with a reduced gain (determined empirically). We consider 100 ions within one peak of a mass spectrum as a good number. We also assume five elements within the particle with equal ionization yield, and each element contributes at least 100 ions to a mass line in the spectrum. This means that we can detect as a lower mass threshold particles providing 500 ions at the ion detector, which corresponds to a particle with at least 1000 atoms. This marks a useful lower particle size and mass detection limit. For a bonding length of approximately 2\,\AA, such a particle would be not larger than 2 or 3\,nm in diameter. However, if a particle contains less than five species, the detection of even smaller particles is not excluded.\\
\\
On the other hand, an impact ionization detector using a multiplier for ion detection can be adjusted to measure larger grains by reducing the detector gain by a factor of 100 or even 1000 by lowering the operating high voltage. This ensures a wide dynamic range to measure mass spectra of nanodust as well as of micron-sized particles with the same instrument electronics. Changes in instrument gain every hour, day or just a month can be foreseen for this reason. This procedure was tested in-flight with \textit{Cassini} CDA.

\section{Conclusions}
Many compelling science questions exist concerning the interaction of ISD (and IDP) with the heliosphere. We highlight the synergies between the two sciences, and what tremendous progress we could make if a dedicated dust suite would fly on an Interstellar Probe to measure dust properties together with the plasma, magnetic field, and pickup ions, during its journey through all regions inside and outside of the heliosphere. The science yield would be increased even more by simultaneous measurements from other missions inside the solar system while ISP is on its journey. The science results may be crucial for understanding the physics and pressure balance of the heliosphere, and the pool of new dust measurements can be used as an extra boundary condition for heliosphere models to help reveal the time-dependent structure and size of the heliosphere. We describe the major advantages for the dust measurements on ISP, including being outside of the heliopause where highly abundant nano-ISD resides, and flying at very high speeds against the flow of ISD – good for detecting dust. From a programmatic point of view, a mission like ISP with a dust detector is crucial, but there is also a need for an optimized long-term monitoring of ISD dynamics parameters (and composition) with broad temporal and spatial coverage in the solar system. The currently existing dust and dust-heliosphere-related instruments each have their advantages and disadvantages for certain types of measurements to help answer the science questions currently posed. With these instruments, we can push forward the boundaries of our knowledge as described here. The topic of dust-heliosphere science is gaining a lot of traction in the community, and collaborations between the continents are important. The ``new space'' launcher industry is expected to allow for instruments with larger detection surfaces in optimized orbits. Finally, solving the science questions presented here will not only benefit dust science and heliosphere science; it will also foster broader synergistic cross-divisional science between heliophysics, astronomy, planetary science and astrobiology, addressing for instance the role of astrospheres in habitability of planetary systems. Such cross-divisional science not only “crosses” the borders of divisions, but also augments science in each of them, thus meeting the exact definition of a true “synergy”.  

\section*{Acknowledgements}
Veerle J.\ Sterken, Lennart R.\ Baalmann and Silvan Hunziker received funding from the European Union’s Horizon 2020 research and innovation programme under grant agreement N$^\circ$~ 851544 \-- ASTRODUST. This work has been carried out within the framework of the NCCR PlanetS supported by the Swiss National Science Foundation under grants 51NF40\_182901 and 51NF40\_205606. Konstantin Herbst acknowledges support from the German Research Foundation priority program SPP 1992 ``Exploring the Diversity of Extrasolar Planets” through the project HE 8392/1-1. The authors would like to thank the anonymous referees for their helpful comments to improve the manuscript. 

\section*{Data Availability}
Data in this paper can be made available upon request.



\bibliographystyle{rasti}
\bibliography{synergies} 

\begin{thebibliography}{149}
\expandafter\ifx\csname natexlab\endcsname\relax\def\natexlab#1{#1}\fi

\bibitem[{Altobelli} et~al.(2016){Altobelli}, {Postberg}, {Fiege}, {Trieloff},
  {Kimura}, {Sterken}, {Hsu}, {Hillier}, {Khawaja}, {Moragas-Klostermeyer},
  {Blum}, {Burton}, {Srama}, {Kempf}, \& {Gruen}]{altobelli:2016}
{Altobelli}, N., {Postberg}, F., {Fiege}, K., {Trieloff}, M., {Kimura}, H.,
  {Sterken}, V.~J., {Hsu}, H.-W., {Hillier}, J., {Khawaja}, N.,
  {Moragas-Klostermeyer}, G., {Blum}, J., {Burton}, M., {Srama}, R., {Kempf},
  S., \& {Gruen}, E., 2016.
\newblock {Flux and composition of interstellar dust at Saturn from Cassini's
  Cosmic Dust Analyzer}, {\it Science\/}, {\bf 352}, 312--318.

\bibitem[{Arai} et~al.(2018){Arai}, {Kobayashi}, {Ishibashi}, {Yoshida},
  {Kimura}, {Wada}, {Senshu}, {Yamada}, {Okudaira}, {Okamoto}, {Kameda},
  {Srama}, {Kruger}, {Ishiguro}, {Yabuta}, {Nakamura}, {Watanabe}, {Ito},
  {Ohtsuka}, {Tachibana}, {Mikouchi}, {Komatsu}, {Nakamura-Messenger},
  {Sasaki}, {Hiroi}, {Abe}, {Urakawa}, {Hirata}, {Demura}, {Komatsu},
  {Noguchi}, {Sekiguchi}, {Inamori}, {Yano}, {Yoshikawa}, {Ohtsubo}, {Okada},
  {Iwata}, {Nishiyama}, {Toyota}, {Kawakatsu}, \& {Takashima}]{arai:2018}
{Arai}, T., {Kobayashi}, M., {Ishibashi}, K., {Yoshida}, F., {Kimura}, H.,
  {Wada}, K., {Senshu}, H., {Yamada}, M., {Okudaira}, O., {Okamoto}, T.,
  {Kameda}, S., {Srama}, R., {Kruger}, H., {Ishiguro}, M., {Yabuta}, H.,
  {Nakamura}, T., {Watanabe}, J., {Ito}, T., {Ohtsuka}, K., {Tachibana}, S.,
  {Mikouchi}, T., {Komatsu}, M., {Nakamura-Messenger}, K., {Sasaki}, S.,
  {Hiroi}, T., {Abe}, S., {Urakawa}, S., {Hirata}, N., {Demura}, H., {Komatsu},
  G., {Noguchi}, T., {Sekiguchi}, T., {Inamori}, T., {Yano}, H., {Yoshikawa},
  M., {Ohtsubo}, T., {Okada}, T., {Iwata}, T., {Nishiyama}, K., {Toyota}, T.,
  {Kawakatsu}, Y., \& {Takashima}, T., 2018.
\newblock {DESTINY+ Mission: Flyby of Geminids Parent Asteroid (3200) Phaethon
  and In-Situ Analyses of Dust Accreting on the Earth}, in {\em 49th Annual
  Lunar and Planetary Science Conference\/}, Lunar and Planetary Science
  Conference, p. 2570.

\bibitem[Auer(1975)]{auer:1975}
Auer, S., 1975.
\newblock Two high resolution velocity vector analyzers for cosmic dust
  particles, {\it Review of Scientific Instruments\/}, {\bf 46}(2), 127--135.

\bibitem[Auer et~al.(2002)Auer, Gr{\"u}n, Srama, Kempf, \& Auer]{auer:2002}
Auer, S., Gr{\"u}n, E., Srama, R., Kempf, S., \& Auer, R., 2002.
\newblock The charge and velocity detector of the cosmic dust analyzer on
  {{Cassini}}, {\it Planetary and Space Science\/}, {\bf 50}(7), 773--779.

\bibitem[Auer et~al.(2008)Auer, Gr{\"u}n, Kempf, Srama, Srowig, Sternovsky, \&
  Tschernjawski]{auer:2008}
Auer, S., Gr{\"u}n, E., Kempf, S., Srama, R., Srowig, A., Sternovsky, Z., \&
  Tschernjawski, V., 2008.
\newblock Characteristics of a dust trajectory sensor, {\it Review of
  Scientific Instruments\/}, {\bf 79}(8), 084501.

\bibitem[{Banks}(1971)]{banks1971}
{Banks}, P.~M., 1971.
\newblock {Interplanetary hydrogen and helium from cosmic dust and the solar
  wind}, {\it \jgr\/}, {\bf 76}(19), 4341.

\bibitem[{Bernardoni} et~al.(2022){Bernardoni}, {Hor{\'a}nyi}, {Doner},
  {Piquette}, {Szalay}, {Poppe}, {James}, {Hunziker}, {Sterken}, {Strub},
  {Olkin}, {Singer}, {Spencer}, {Stern}, \& {Weaver}]{bernardoni:2022}
{Bernardoni}, E., {Hor{\'a}nyi}, M., {Doner}, A., {Piquette}, M., {Szalay},
  J.~R., {Poppe}, A.~R., {James}, D., {Hunziker}, S., {Sterken}, V., {Strub},
  P., {Olkin}, C., {Singer}, K.~N., {Spencer}, J., {Stern}, A., \& {Weaver},
  H., 2022.
\newblock {Student Dust Counter Status Report: The First 50 au}, {\it The
  Planetary Science Journal\/}, {\bf 3}(3), 69.

\bibitem[{Bradley}(2014)]{bradley:2014}
{Bradley}, J.~P., 2014.
\newblock {Early Solar Nebula Grains - Interplanetary Dust Particles}, in {\em
  Meteorites and Cosmochemical Processes\/}, vol.~1, pp. 287--308, ed. {Davis},
  A.~M.

\bibitem[{Brandt} et~al.(2022){Brandt}, {Provornikova}, {Cocoros}, {Turner},
  {DeMajistre}, {Runyon}, {Lisse}, {Bale}, {Kurth}, {Galli}, {Wurz}, {McNutt},
  {Wimmer-Schweingruber}, {Linsky}, {Redfield}, {Kollmann}, {Mandt}, {Rymer},
  {Roelof}, {Kinnison}, {Opher}, {Hill}, \& {Paul}]{brandt:2022}
{Brandt}, P.~C., {Provornikova}, E.~A., {Cocoros}, A., {Turner}, D.,
  {DeMajistre}, R., {Runyon}, K., {Lisse}, C.~M., {Bale}, S., {Kurth}, W.~S.,
  {Galli}, A., {Wurz}, P., {McNutt}, R.~L., {Wimmer-Schweingruber}, R.,
  {Linsky}, J., {Redfield}, S., {Kollmann}, P., {Mandt}, K.~E., {Rymer}, A.~M.,
  {Roelof}, E.~C., {Kinnison}, J., {Opher}, M., {Hill}, M.~E., \& {Paul},
  M.~V., 2022.
\newblock {Interstellar Probe: Humanity's exploration of the Galaxy Begins},
  {\it Acta Astronautica\/}, {\bf 199}, 364--373.

\bibitem[{Brandt} et~al.(2023){Brandt}, {Provornikova}, {Bale}, {Cocoros},
  {DeMajistre}, {Dialynas}, {Elliott}, {Eriksson}, {Fields}, {Galli}, {Hill},
  {Horanyi}, {Horbury}, {Hunziker}, {Kollmann}, {Kinnison}, {Fountain},
  {Krimigis}, {Kurth}, {Linsky}, {Lisse}, {Mandt}, {Magnes}, {McNutt},
  {Miller}, {Moebius}, {Mostafavi}, {Opher}, {Paxton}, {Plaschke}, {Poppe},
  {Roelof}, {Runyon}, {Redfield}, {Schwadron}, {Sterken}, {Swaczyna}, {Szalay},
  {Turner}, {Vannier}, {Wimmer-Schweingruber}, {Wurz}, \&
  {Zirnstein}]{brandt_ea:2023}
{Brandt}, P.~C., {Provornikova}, E., {Bale}, S.~D., {Cocoros}, A.,
  {DeMajistre}, R., {Dialynas}, K., {Elliott}, H.~A., {Eriksson}, S., {Fields},
  B., {Galli}, A., {Hill}, M.~E., {Horanyi}, M., {Horbury}, T., {Hunziker}, S.,
  {Kollmann}, P., {Kinnison}, J., {Fountain}, G., {Krimigis}, S.~M., {Kurth},
  W.~S., {Linsky}, J., {Lisse}, C.~M., {Mandt}, K.~E., {Magnes}, W., {McNutt},
  R.~L., {Miller}, J., {Moebius}, E., {Mostafavi}, P., {Opher}, M., {Paxton},
  L., {Plaschke}, F., {Poppe}, A.~R., {Roelof}, E.~C., {Runyon}, K.,
  {Redfield}, S., {Schwadron}, N., {Sterken}, V., {Swaczyna}, P., {Szalay}, J.,
  {Turner}, D., {Vannier}, H., {Wimmer-Schweingruber}, R., {Wurz}, P., \&
  {Zirnstein}, E.~J., 2023.
\newblock {Future Exploration of the Outer Heliosphere and Very Local
  Interstellar Medium by Interstellar Probe}, {\it \ssr\/}, {\bf 219}(2), 18.

\bibitem[{Brown} \& {Borovi{\v{c}}ka}(2023)]{brown:2023}
{Brown}, P.~G. \& {Borovi{\v{c}}ka}, J., 2023.
\newblock {On the Proposed Interstellar Origin of the USG 20140108 Fireball},
  {\it arXiv e-prints\/}, p. arXiv:2306.14267.

\bibitem[{Burchell} et~al.(1999){Burchell}, {Cole}, {McDonnell}, \&
  {Zarnecki}]{burchell:1999}
{Burchell}, M.~J., {Cole}, M.~J., {McDonnell}, J.~A.~M., \& {Zarnecki}, J.~C.,
  1999.
\newblock {Hypervelocity impact studies using the 2 MV Van de Graaff
  accelerator and two-stage light gas gun of the University of Kent at
  Canterbury}, {\it Measurement Science and Technology\/}, {\bf 10}(1), 41--50.

\bibitem[{Carpenter} et~al.(2007){Carpenter}, {Stevenson}, {Fraser}, {Bridges},
  {Kearsley}, {Chater}, \& {Hainsworth}]{carpenter:2007}
{Carpenter}, J.~D., {Stevenson}, T.~J., {Fraser}, G.~W., {Bridges}, J.~C.,
  {Kearsley}, A.~T., {Chater}, R.~J., \& {Hainsworth}, S.~V., 2007.
\newblock {Nanometer hypervelocity dust impacts in low Earth orbit}, {\it
  Journal of Geophysical Research (Planets)\/}, {\bf 112}(E8), E08008.

\bibitem[{Cummings} et~al.(2002){Cummings}, {Stone}, \&
  {Steenberg}]{cummings+2002}
{Cummings}, A.~C., {Stone}, E.~C., \& {Steenberg}, C.~D., 2002.
\newblock {Erratum: ``Composition of Anomalous Cosmic Rays and Other
  Heliospheric Ions`` (<A href=''/abs/2002ApJ...578..194C''>ApJ, 578, 194
  [2002]</A>)}, {\it \apj\/}, {\bf 581}(2), 1413--1413.

\bibitem[{Darling}(2019)]{darling:2019}
{Darling}, S., 2019.
\newblock 25 years of science in the solar wind,
  \url{https://www.nasa.gov/feature/goddard/2019/25-years-of-science-in-the-solar-wind},
  Accessed: 2022-11-01.

\bibitem[{Decker} et~al.(2005){Decker}, {Krimigis}, {Roelof}, {Hill},
  {Armstrong}, {Gloeckler}, {Hamilton}, \& {Lanzerotti}]{Decker_ea:2005}
{Decker}, R.~B., {Krimigis}, S.~M., {Roelof}, E.~C., {Hill}, M.~E.,
  {Armstrong}, T.~P., {Gloeckler}, G., {Hamilton}, D.~C., \& {Lanzerotti},
  L.~J., 2005.
\newblock {Voyager 1 in the Foreshock, Termination Shock, and Heliosheath},
  {\it Science\/}, {\bf 309}(5743), 2020--2024.

\bibitem[{Decker} et~al.(2008){Decker}, {Krimigis}, {Roelof}, {Hill},
  {Armstrong}, {Gloeckler}, {Hamilton}, \& {Lanzerotti}]{decker_ea:2008}
{Decker}, R.~B., {Krimigis}, S.~M., {Roelof}, E.~C., {Hill}, M.~E.,
  {Armstrong}, T.~P., {Gloeckler}, G., {Hamilton}, D.~C., \& {Lanzerotti},
  L.~J., 2008.
\newblock {Mediation of the solar wind termination shock by non-thermal ions},
  {\it \nat\/}, {\bf 454}(7200), 67--70.

\bibitem[{Dialynas} et~al.(2019){Dialynas}, {Krimigis}, {Decker}, \&
  {Mitchell}]{dialynas_ea:2019}
{Dialynas}, K., {Krimigis}, S.~M., {Decker}, R.~B., \& {Mitchell}, D.~G., 2019.
\newblock {Plasma Pressures in the Heliosheath From Cassini ENA and Voyager 2
  Measurements: Validation by the Voyager 2 Heliopause Crossing}, {\it \grl\/},
  {\bf 46}(14), 7911--7919.

\bibitem[{Dialynas} et~al.(2020){Dialynas}, {Galli}, {Dayeh}, {Cummings},
  {Decker}, {Fuselier}, {Gkioulidou}, {Roussos}, {Krimigis}, {Mitchell},
  {Richardson}, \& {Opher}]{dialynas_ea:2020}
{Dialynas}, K., {Galli}, A., {Dayeh}, M.~A., {Cummings}, A.~C., {Decker},
  R.~B., {Fuselier}, S.~A., {Gkioulidou}, M., {Roussos}, E., {Krimigis}, S.~M.,
  {Mitchell}, D.~G., {Richardson}, J.~D., \& {Opher}, M., 2020.
\newblock {Combined {\ensuremath{\sim}}10 eV to {\ensuremath{\sim}}344 MeV
  Particle Spectra and Pressures in the Heliosheath along the Voyager 2
  Trajectory}, {\it \apjl\/}, {\bf 905}(2), L24.

\bibitem[{Dialynas} et~al.(2022){Dialynas}, {Krimigis}, {Decker}, {Hill},
  {Mitchell}, {Hsieh}, {Hilchenbach}, \& {Czechowski}]{dialynas_ea:2022}
{Dialynas}, K., {Krimigis}, S.~M., {Decker}, R.~B., {Hill}, M., {Mitchell},
  D.~G., {Hsieh}, K.~C., {Hilchenbach}, M., \& {Czechowski}, A., 2022.
\newblock {The Structure of the Global Heliosphere as Seen by In-Situ Ions from
  the Voyagers and Remotely Sensed ENAs from Cassini}, {\it \ssr\/}, {\bf
  218}(4), 21.

\bibitem[{Dialynas} et~al.(2023){Dialynas}, {Sterken}, {Brandt}, {Burlaga},
  {Berdichevsky}, {Decker}, {Della Torre}, {DeMajistre}, {Galli}, {Gkioulidou},
  {Hill}, {Krimigis}, {Kornbleuth}, {Kurth}, {Lavraud}, {McNutt}, {Mitchell},
  {Mostafavi}, {Nikoukar}, {Opher}, {Provornikova}, {Roelof}, {Rancoita},
  {Richardson}, {Roussos}, {Sok{\'o}{\l}}, {La Vacca}, {Westlake}, \&
  {Chen}]{dialynas_ea:2023}
{Dialynas}, K., {Sterken}, V.~J., {Brandt}, P.~C., {Burlaga}, L.,
  {Berdichevsky}, D.~B., {Decker}, R.~B., {Della Torre}, S., {DeMajistre}, R.,
  {Galli}, A., {Gkioulidou}, M., {Hill}, M.~E., {Krimigis}, S.~M.,
  {Kornbleuth}, M., {Kurth}, W., {Lavraud}, B., {McNutt}, R., {Mitchell},
  D.~G., {Mostafavi}, P.~S., {Nikoukar}, R., {Opher}, M., {Provornikova}, E.,
  {Roelof}, E.~C., {Rancoita}, P.~G., {Richardson}, J.~D., {Roussos}, E.,
  {Sok{\'o}{\l}}, J.~M., {La Vacca}, G., {Westlake}, J., \& {Chen}, T.~Y.,
  2023.
\newblock {A future interstellar probe on the dynamic heliosphere and its
  interaction with the very local interstellar medium: In-situ particle and
  fields measurements and remotely sensed ENAs}, {\it Frontiers in Astronomy
  and Space Sciences\/}, {\bf 10}, 1061969.

\bibitem[{Draine}(2003)]{draine:2003}
{Draine}, B.~T., 2003.
\newblock {Interstellar Dust Grains}, {\it \araa\/}, {\bf 41}, 241--289.

\bibitem[{Draine}(2009)]{draine:2009}
{Draine}, B.~T., 2009.
\newblock {Perspectives on Interstellar Dust Inside and Outside of the
  Heliosphere}, {\it Space~Sci.~Rev.\/}, {\bf 143}(1-4), 333--345.

\bibitem[{Draine} \& {Li}(2007)]{Draine2007}
{Draine}, B.~T. \& {Li}, A., 2007.
\newblock {Infrared Emission from Interstellar Dust. IV. The
  Silicate-Graphite-PAH Model in the Post-Spitzer Era}, {\it \apj\/}, {\bf
  657}(2), 810--837.

\bibitem[{Friichtenicht} \& {Slattery}(1963)]{friichtenicht:1963}
{Friichtenicht}, J.~F. \& {Slattery}, J.~C., 1963.
\newblock {Ionization associated with hypervelocity impact}, Tech. rep., NASA.

\bibitem[{Frisch} et~al.(2022){Frisch}, {Piirola}, {Berdyugin}, {Heiles},
  {Cole}, {Hill}, {Magalh{\~a}es}, {Wiktorowicz}, {Bailey}, {Cotton},
  {Kedziora-Chudczer}, {Schwadron}, {Bzowski}, {McComas}, {Zirnstein},
  {Funsten}, {Harlingten}, \& {Redfield}]{frisch:2022}
{Frisch}, P.~C., {Piirola}, V., {Berdyugin}, A.~B., {Heiles}, C., {Cole}, A.,
  {Hill}, K., {Magalh{\~a}es}, A.~M., {Wiktorowicz}, S.~J., {Bailey}, J.,
  {Cotton}, D.~V., {Kedziora-Chudczer}, L., {Schwadron}, N.~A., {Bzowski}, M.,
  {McComas}, D.~J., {Zirnstein}, E.~J., {Funsten}, H.~O., {Harlingten}, C., \&
  {Redfield}, S., 2022.
\newblock {Whence the Interstellar Magnetic Field Shaping the Heliosphere?},
  {\it ApJs\/}, {\bf 259}(2), 48.

\bibitem[Funase et~al.(2020)Funase, Ikari, Miyoshi, Kawabata, Nakajima, Nomura,
  Funabiki, Ishikawa, Kakihara, Matsushita, Takahashi, Yanagida, Mori, Murata,
  Shibukawa, Suzumoto, Fujiwara, Tomita, Aohama, Iiyama, Ishiwata, Kondo,
  Mikuriya, Seki, Koizumi, Asakawa, Nishii, Hattori, Saito, Kikuchi, Kobayashi,
  Tomiki, Torii, Ito, Campagnola, Ozaki, Baresi, Yoshikawa, Yoshioka, Kuwabara,
  Hikida, Arao, Abe, Yanagisawa, Fuse, Masuda, Yano, Hirai, Arai, Jitsukawa,
  Ishioka, Nakano, Ikenaga, \& Hashimoto]{funase:2020}
Funase, R., Ikari, S., Miyoshi, K., Kawabata, Y., Nakajima, S., Nomura, S.,
  Funabiki, N., Ishikawa, A., Kakihara, K., Matsushita, S., Takahashi, R.,
  Yanagida, K., Mori, D., Murata, Y., Shibukawa, T., Suzumoto, R., Fujiwara,
  M., Tomita, K., Aohama, H., Iiyama, K., Ishiwata, S., Kondo, H., Mikuriya,
  W., Seki, H., Koizumi, H., Asakawa, J., Nishii, K., Hattori, A., Saito, Y.,
  Kikuchi, K., Kobayashi, Y., Tomiki, A., Torii, W., Ito, T., Campagnola, S.,
  Ozaki, N., Baresi, N., Yoshikawa, I., Yoshioka, K., Kuwabara, M., Hikida, R.,
  Arao, S., Abe, S., Yanagisawa, M., Fuse, R., Masuda, Y., Yano, H., Hirai, T.,
  Arai, K., Jitsukawa, R., Ishioka, E., Nakano, H., Ikenaga, T., \& Hashimoto,
  T., 2020.
\newblock Mission to {{Earth}}\textendash{{Moon Lagrange Point}} by a {{6U
  CubeSat}}: {{EQUULEUS}}, {\it IEEE Aerospace and Electronic Systems
  Magazine\/}, {\bf 35}(3), 30--44.

\bibitem[Föhn et~al.(2021)Föhn, Galli, Vorburger, Tulej, Lasi, Riedo, Fausch,
  Althaus, Brüngger, Fahrer, Gerber, Lüthi, Munz, Oeschger, Piazza, \&
  Wurz]{foehn21}
Föhn, M., Galli, A., Vorburger, A., Tulej, M., Lasi, D., Riedo, A., Fausch,
  R.~G., Althaus, M., Brüngger, S., Fahrer, P., Gerber, M., Lüthi, M., Munz,
  H.~P., Oeschger, S., Piazza, D., \& Wurz, P., 2021.
\newblock Description of the mass spectrometer for the jupiter icy moons
  explorer mission, in {\em 2021 IEEE Aerospace Conference (50100)\/}, pp.
  1--14.

\bibitem[{Geiss} et~al.(1995){Geiss}, {Gloeckler}, {Fisk}, \& {von
  Steiger}]{geiss+1995}
{Geiss}, J., {Gloeckler}, G., {Fisk}, L.~A., \& {von Steiger}, R., 1995.
\newblock {C$^{+}$ pickup ions in the heliosphere and their origin}, {\it
  \jgr\/}, {\bf 100}(A12), 23373--23378.

\bibitem[{Gloeckler} et~al.(2000){Gloeckler}, {Fisk}, {Geiss}, {Schwadron}, \&
  {Zurbuchen}]{gloeckler+2000}
{Gloeckler}, G., {Fisk}, L.~A., {Geiss}, J., {Schwadron}, N.~A., \&
  {Zurbuchen}, T.~H., 2000.
\newblock {Elemental composition of the inner source pickup ions}, {\it
  \jgr\/}, {\bf 105}(A4), 7459--7464.

\bibitem[G{\"o}ller \& Gr{\"u}n(1989)]{goller:1989}
G{\"o}ller, J. \& Gr{\"u}n, E., 1989.
\newblock Calibration of the {{Galileo}}/{{Ulysses}} dust detectors with
  different projectile materials and at varying impact angles, {\it Planetary
  and Space Science\/}, {\bf 37}(10), 1197--1206.

\bibitem[{Gregg} \& {Wiegert}(2023)]{gregg:2023}
{Gregg}, C.~R. \& {Wiegert}, P., 2023.
\newblock {The development of interstellar meteoroid streams}, in {\em
  Asteroids, Comets, Meteors 2023\/}, p. 2164/2851.

\bibitem[{Gr{\"u}n}(1984)]{gruen:1984}
{Gr{\"u}n}, E., 1984.
\newblock {Impact ionization from gold, aluminum and PCB-Z.}, in {\em ESA
  Special Publication\/}, vol. 224 of {\bf ESA Special Publication}, pp.
  39--41.

\bibitem[{Gr{\"u}n}(1993)]{gruen:1993b}
{Gr{\"u}n}, E., 1993.
\newblock {Dust in the planetary system}, {\it Advances in Space Research\/},
  {\bf 13}(10), 139--151.

\bibitem[{Gr{\"u}n} \& {Landgraf}(2000)]{gruen:2000}
{Gr{\"u}n}, E. \& {Landgraf}, M., 2000.
\newblock {Collisional consequences of big interstellar grains}, {\it \jgr\/},
  {\bf 105}(A5), 10291--10298.

\bibitem[{Gr{\"u}n} \& {Svestka}(1996)]{gruen:1996}
{Gr{\"u}n}, E. \& {Svestka}, J., 1996.
\newblock {Physics of Interplanetary and Interstellar Dust}, {\it \ssr\/}, {\bf
  78}(1-2), 347--360.

\bibitem[Gr{\"u}n et~al.(1992)Gr{\"u}n, Fechtig, Hanner, Kissel, Lindblad,
  Linkert, Maas, Morfill, \& Zook]{grun:1992}
Gr{\"u}n, E., Fechtig, H., Hanner, M.~S., Kissel, J., Lindblad, B.-A., Linkert,
  D., Maas, D., Morfill, G.~E., \& Zook, H.~A., 1992.
\newblock The {{Galileo Dust Detector}}, {\it Space Science Reviews\/}, {\bf
  60}(1), 317--340.

\bibitem[Gr{\"u}n et~al.(1993)Gr{\"u}n, Zook, Baguhl, Balogh, Bame, Fechtig,
  Forsyth, Hanner, Horanyi, Kissel, Lindblad, Linkert, Linkert, Mann, McDonnel,
  Morfill, Phillips, Polanskey, Schwehm, Siddique, Staubach, Svestka, \&
  Taylor]{gruen:1993}
Gr{\"u}n, E., Zook, H., Baguhl, M., Balogh, A., Bame, S., Fechtig, H., Forsyth,
  R., Hanner, M., Horanyi, M., Kissel, J., Lindblad, B.-A., Linkert, D.,
  Linkert, G., Mann, I., McDonnel, J., Morfill, G., Phillips, J., Polanskey,
  C., Schwehm, G., Siddique, N., Staubach, P., Svestka, J., \& Taylor, A.,
  1993.
\newblock Discovery of {Jovian} dust streams and interstellar grains by the
  {Ulysses} spacecraft, {\it Nature\/}, {\bf 362}, 428--430.

\bibitem[Gr{\"u}n et~al.(2005)Gr{\"u}n, Srama, Kr{\"u}ger, Kempf, Dikarev,
  Helfert, \& {Moragas-Klostermeyer}]{grun:2005}
Gr{\"u}n, E., Srama, R., Kr{\"u}ger, H., Kempf, S., Dikarev, V., Helfert, S.,
  \& {Moragas-Klostermeyer}, G., 2005.
\newblock 2002 {{Kuiper}} prize lecture: {{Dust Astronomy}}, {\it Icarus\/},
  {\bf 174}(1), 1--14.

\bibitem[{Gruntman} \& {Izmodenov}(2004)]{gruntman+2004}
{Gruntman}, M. \& {Izmodenov}, V., 2004.
\newblock {Mass transport in the heliosphere by energetic neutral atoms}, {\it
  Journal of Geophysical Research (Space Physics)\/}, {\bf 109}(A12), A12108.

\bibitem[{Gurnett} et~al.(1983){Gurnett}, {Grun}, {Gallagher}, {Kurth}, \&
  {Scarf}]{gurnett:1983}
{Gurnett}, D.~A., {Grun}, E., {Gallagher}, D., {Kurth}, W.~S., \& {Scarf},
  F.~L., 1983.
\newblock {Micron-sized particles detected near Saturn by the Voyager plasma
  wave instrument}, {\it Icarus\/}, {\bf 53}, 236--254.

\bibitem[{Hajdukov{\'a}} et~al.(2019){Hajdukov{\'a}}, {Sterken}, \&
  {Wiegert}]{hajdukova:2019}
{Hajdukov{\'a}}, M., J., {Sterken}, V., \& {Wiegert}, P., 2019.
\newblock {Interstellar Meteoroids}, in {\em Meteoroids: Sources of Meteors on
  Earth and Beyond\/}, p. 235, eds {Ryabova}, G.~O., {Asher}, D.~J., \&
  {Campbell-Brown}, M.~J.

\bibitem[{Hajdukova} et~al.(2020){Hajdukova}, {Sterken}, {Wiegert}, \&
  {Korno{\v{s}}}]{hajdukova:2020}
{Hajdukova}, M., {Sterken}, V., {Wiegert}, P., \& {Korno{\v{s}}}, L., 2020.
\newblock {The challenge of identifying interstellar meteors}, {\it \planss\/},
  {\bf 192}, 105060.

\bibitem[{Hajdukova} et~al.(2023){Hajdukova}, {Stober}, {Barghini},
  {Vaubaillon}, {Sterken}, {Durisova}, \& {Koten}]{hajdukova:2023}
{Hajdukova}, M., {Stober}, G., {Barghini}, D., {Vaubaillon}, J., {Sterken}, V.,
  {Durisova}, S., \& {Koten}, P., 2023.
\newblock {No evidence for interstellar meteors in the CNEOS database}, {\it in
  prep.\/}.

\bibitem[{Heck} et~al.(2012){Heck}, {Hoppe}, \& {Huth}]{heck:2012}
{Heck}, P.~R., {Hoppe}, P., \& {Huth}, J., 2012.
\newblock {Sulfur four isotope NanoSIMS analysis of comet-81P/Wild 2 dust in
  impact craters on aluminum foil C2037N from NASA's Stardust mission}, {\it
  Meteorit.\ Planet.\ Sci.\/}, {\bf 47}(4), 649--659.

\bibitem[{Hervig} et~al.(2022){Hervig}, {Malaspina}, {Sterken}, {Wilson}~III,
  {Hunziker}, \& {Bailey}]{hervig:2022}
{Hervig}, M.~E., {Malaspina}, D., {Sterken}, V.~J., {Wilson}~III, L.~B.,
  {Hunziker}, S., \& {Bailey}, S.~M., 2022.
\newblock {Decadal and annual variations in meteoric flux from Ulysses, Wind
  and SOFIE observations. Accepted}, {\it \jgr\/}.

\bibitem[Hillier et~al.(2014)Hillier, Sternovsky, Armes, Fielding, Postberg,
  Bugiel, Drake, Srama, Kearsley, \& Trieloff]{hillier:2014}
Hillier, J.~K., Sternovsky, Z., Armes, S.~P., Fielding, L.~A., Postberg, F.,
  Bugiel, S., Drake, K., Srama, R., Kearsley, A.~T., \& Trieloff, M., 2014.
\newblock Impact ionisation mass spectrometry of polypyrrole-coated pyrrhotite
  microparticles, {\it Planetary and Space Science\/}, {\bf 97}, 9--22.

\bibitem[Hillier et~al.(2018)Hillier, Sternovsky, Kempf, Trieloff, Guglielmino,
  Postberg, \& Price]{hillier:2018}
Hillier, J.~K., Sternovsky, Z., Kempf, S., Trieloff, M., Guglielmino, M.,
  Postberg, F., \& Price, M.~C., 2018.
\newblock Impact ionisation mass spectrometry of platinum-coated olivine and
  magnesite-dominated cosmic dust analogues, {\it Planetary and Space
  Science\/}, {\bf 156}, 96--110.

\bibitem[Hirai et~al.(2014)Hirai, Cole, Fujii, Hasegawa, Iwai, Kobayashi,
  Srama, \& Yano]{hirai:2014}
Hirai, T., Cole, M.~J., Fujii, M., Hasegawa, S., Iwai, T., Kobayashi, M.,
  Srama, R., \& Yano, H., 2014.
\newblock Microparticle impact calibration of the {{Arrayed Large-Area Dust
  Detectors}} in {{INterplanetary}} space ({{ALADDIN}}) onboard the solar power
  sail demonstrator {{IKAROS}}, {\it Planetary and Space Science\/}, {\bf 100},
  87--97.

\bibitem[{Hony} et~al.(2002){Hony}, {Waters}, \& {Tielens}]{hony:2002}
{Hony}, S., {Waters}, L.~B.~F.~M., \& {Tielens}, A.~G.~G.~M., 2002.
\newblock {The carrier of the ``30'' mu m emission feature in evolved stars. A
  simple model using magnesium sulfide}, {\it \aap\/}, {\bf 390}, 533--553.

\bibitem[{Hoppe} et~al.(2012){Hoppe}, {Fujiya}, \& {Zinner}]{hoppe:2012}
{Hoppe}, P., {Fujiya}, W., \& {Zinner}, E., 2012.
\newblock {Sulfur Molecule Chemistry in Supernova Ejecta Recorded by Silicon
  Carbide Stardust}, {\it \apjl\/}, {\bf 745}(2), L26.

\bibitem[{Hoppe} et~al.(2013){Hoppe}, {Cohen}, \& {Meibom}]{hoppe:2013}
{Hoppe}, P., {Cohen}, S., \& {Meibom}, A., 2013.
\newblock {NanoSIMS: Technical Aspects and Applications in Cosmochemistry and
  Biological Geochemistry}, {\it Geostandards and Geoanalytical Research\/},
  {\bf 37}(2), 111--154.

\bibitem[{Hoppe} et~al.(2017){Hoppe}, {Leitner}, \&
  {Kodol{\'a}nyi}]{hoppe:2017}
{Hoppe}, P., {Leitner}, J., \& {Kodol{\'a}nyi}, J., 2017.
\newblock {The stardust abundance in the local interstellar cloud at the birth
  of the Solar System}, {\it Nature Astronomy\/}, {\bf 1}, 617--620.

\bibitem[{Hoppe} et~al.(2021){Hoppe}, {Leitner}, {Kodol{\'a}nyi}, \&
  {Vollmer}]{hoppe:2021}
{Hoppe}, P., {Leitner}, J., {Kodol{\'a}nyi}, J., \& {Vollmer}, C., 2021.
\newblock {Isotope Systematics of Presolar Silicate Grains: New Insights from
  Magnesium and Silicon}, {\it \apj\/}, {\bf 913}(1), 10.

\bibitem[{Hor{\'a}nyi}(2000)]{horanyi:2000}
{Hor{\'a}nyi}, M., 2000.
\newblock {Dust streams from Jupiter and Saturn}, {\it Physics of Plasmas\/},
  {\bf 7}(10), 3847--3850.

\bibitem[Hor{\'a}nyi et~al.(2008)Hor{\'a}nyi, Hoxie, James, Poppe, Bryant,
  Grogan, Lamprecht, Mack, Bagenal, Batiste, Bunch, Chanthawanich, Christensen,
  Colgan, Dunn, Drake, Fernandez, Finley, Holland, Jenkins, Krauss, Krauss,
  Krauss, Lankton, Mitchell, Neeland, Reese, Rash, Tate, Vaudrin, \&
  Westfall]{horanyi:2008}
Hor{\'a}nyi, M., Hoxie, V., James, D., Poppe, A., Bryant, C., Grogan, B.,
  Lamprecht, B., Mack, J., Bagenal, F., Batiste, S., Bunch, N., Chanthawanich,
  T., Christensen, F., Colgan, M., Dunn, T., Drake, G., Fernandez, A., Finley,
  T., Holland, G., Jenkins, A., Krauss, C., Krauss, E., Krauss, O., Lankton,
  M., Mitchell, C., Neeland, M., Reese, T., Rash, K., Tate, G., Vaudrin, C., \&
  Westfall, J., 2008.
\newblock The {{Student Dust Counter}} on the {{New Horizons Mission}}, {\it
  Space Science Reviews\/}, {\bf 140}(1), 387--402.

\bibitem[Hor{\'a}nyi et~al.(2019)Hor{\'a}nyi, Kempf, Sternovsky, Tucker,
  Pokorn{\'y}, Turner, {Castillo-Rogez}, B{\'a}lint, West, \&
  Szalay]{horanyi:2019}
Hor{\'a}nyi, M., Kempf, S., Sternovsky, Z., Tucker, S., Pokorn{\'y}, P.,
  Turner, N.~J., {Castillo-Rogez}, J.~C., B{\'a}lint, T., West, J.~L., \&
  Szalay, J.~R., 2019.
\newblock Fragments from the {{Origins}} of the {{Solar System}} and our
  {{Interstellar Locale}} ({{FOSSIL}}): {{A Cometary}}, {{Asteroidal}}, and
  {{Interstellar Dust Mission Concept}}, in {\em 2019 {{IEEE Aerospace
  Conference}}\/}, pp. 1--12.

\bibitem[{Hsu} et~al.(2011){Hsu}, {Postberg}, {Kempf}, {Trieloff}, {Burton},
  {Roy}, {Moragas-Klostermeyer}, \& {Srama}]{hsu:2011}
{Hsu}, H.~W., {Postberg}, F., {Kempf}, S., {Trieloff}, M., {Burton}, M., {Roy},
  M., {Moragas-Klostermeyer}, G., \& {Srama}, R., 2011.
\newblock {Stream particles as the probe of the dust-plasma-magnetosphere
  interaction at Saturn}, {\it Journal of Geophysical Research (Space
  Physics)\/}, {\bf 116}(A9), A09215.

\bibitem[{Hsu} et~al.(2018){Hsu}, {Schmidt}, {Kempf}, {Postberg},
  {Moragas-Klostermeyer}, {Sei{\ss}}, {Hoffmann}, {Burton}, {Ye}, {Kurth},
  {Hor{\'a}nyi}, {Khawaja}, {Spahn}, {Schirdewahn}, {O'Donoghue}, {Moore},
  {Cuzzi}, {Jones}, \& {Srama}]{hsu:2018}
{Hsu}, H.-W., {Schmidt}, J., {Kempf}, S., {Postberg}, F.,
  {Moragas-Klostermeyer}, G., {Sei{\ss}}, M., {Hoffmann}, H., {Burton}, M.,
  {Ye}, S., {Kurth}, W.~S., {Hor{\'a}nyi}, M., {Khawaja}, N., {Spahn}, F.,
  {Schirdewahn}, D., {O'Donoghue}, J., {Moore}, L., {Cuzzi}, J., {Jones},
  G.~H., \& {Srama}, R., 2018.
\newblock {In situ collection of dust grains falling from Saturn's rings into
  its atmosphere}, {\it Science\/}, {\bf 362}(6410), aat3185.

\bibitem[{Hsu} et~al.(2022){Hsu}, {Poppe}, {Szalay}, {Horanyi}, {Sterken},
  {Chen}, \& {Mann}]{hsu:2022}
{Hsu}, H.-W., {Poppe}, A., {Szalay}, J., {Horanyi}, M., {Sterken}, V., {Chen},
  T.~Y., \& {Mann}, I., 2022.
\newblock {Science opportunities enabled by in situ cosmic dust detection
  technology for heliophysics and beyond (White paper submitted to for the
  decadal survey in Solar and Space Physics (Heliophysics) 2024-2033)}, {\it
  Bulletin of the AAS (BAAS)\/}.

\bibitem[{Hunziker} et~al.(2022){Hunziker}, {Moragas-Klostermeyer}, {Hillier},
  {Fielding}, {Hornung}, {Lovett}, {Armes}, {Fontanese}, {James}, {Hsu},
  {Herrmann}, {Fechler}, {Poch}, {Pommerol}, {Srama}, {Malaspina}, \&
  {Sterken}]{hunziker:2022}
{Hunziker}, S., {Moragas-Klostermeyer}, G., {Hillier}, J.~K., {Fielding},
  L.~A., {Hornung}, K., {Lovett}, J.~R., {Armes}, S.~P., {Fontanese}, J.,
  {James}, D., {Hsu}, H.~W., {Herrmann}, I., {Fechler}, N., {Poch}, O.,
  {Pommerol}, A., {Srama}, R., {Malaspina}, D., \& {Sterken}, V.~J., 2022.
\newblock {Impact ionization dust detection with compact, hollow and fluffy
  dust analogs}, {\it \planss\/}, {\bf 220}, 105536.

\bibitem[{Hunziker} et~al.(2023){Hunziker}, {Strub}, {Brandt}, {Kr\"{u}ger},
  {Janisch}, {Hsu}, {Postberg}, {Horanyi}, {Szalay}, {Poppe}, {Lisse}, \&
  {Sterken}]{hunziker:2023isp}
{Hunziker}, S., {Strub}, P., {Brandt}, P., {Kr\"{u}ger}, H., {Janisch}, T.,
  {Hsu}, H.-W., {Postberg}, F., {Horanyi}, M., {Szalay}, J., {Poppe}, A.,
  {Lisse}, C., \& {Sterken}, V.~J., 2023.
\newblock {Interstellar Probe: a goldmine for interstellar dust research}, {\it
  in prep\/}.

\bibitem[{Interstellar Probe}(2023)]{isp:website}
{Interstellar Probe}, 2023.
\newblock {Interstellar Probe -- Science},
  \url{https://interstellarprobe.jhuapl.edu/Science/}, Accessed on 2023-06-30.

\bibitem[James et~al.(2010)James, Hoxie, \& Horanyi]{james:2010}
James, D., Hoxie, V., \& Horanyi, M., 2010.
\newblock Polyvinylidene fluoride dust detector response to particle impacts,
  {\it Review of Scientific Instruments\/}, {\bf 81}(3), 034501.

\bibitem[Jaynes(2023)]{jaynes:2023}
Jaynes, A.~N., 2023.
\newblock personal communication.

\bibitem[{Kallenbach} et~al.(2000){Kallenbach}, {Geiss}, {Gloeckler}, \& {von
  Steiger}]{kallenbach+2000}
{Kallenbach}, R., {Geiss}, J., {Gloeckler}, G., \& {von Steiger}, R., 2000.
\newblock {Pick-up Ion Measurements in the Heliosphere - A Review}, {\it
  \apss\/}, {\bf 274}, 97--114.

\bibitem[{Keller} \& {Flynn}(2022)]{keller:2022}
{Keller}, L.~P. \& {Flynn}, G.~J., 2022.
\newblock {Evidence for a significant Kuiper belt dust contribution to the
  zodiacal cloud}, {\it Nature Astronomy\/}, {\bf 6}, 731--735.

\bibitem[{Keller} \& {Rahman}(2011)]{keller:2011}
{Keller}, L.~P. \& {Rahman}, Z., 2011.
\newblock {Irradiation of FeS: Relative Sputtering Rates of Troilite and Mg
  Silicates}, {\it Meteoritics and Planetary Science Supplement\/}, {\bf 74},
  5455.

\bibitem[Kempf(2018)]{kempf:2018}
Kempf, S., 2018.
\newblock The {{Surface Dust Analyzer}} ({{SUDA}}) on {{Europa Clipper}}, in
  {\em European {{Planetary Science Congress}} 2018\/}, pp. EPSC2018--462.

\bibitem[{Kempf} et~al.(2005){Kempf}, {Srama}, {Hor{\'a}nyi}, {Burton},
  {Helfert}, {Moragas-Klostermeyer}, {Roy}, \& {Gr{\"u}n}]{kempf:2005}
{Kempf}, S., {Srama}, R., {Hor{\'a}nyi}, M., {Burton}, M., {Helfert}, S.,
  {Moragas-Klostermeyer}, G., {Roy}, M., \& {Gr{\"u}n}, E., 2005.
\newblock {High-velocity streams of dust originating from Saturn}, {\it
  \nat\/}, {\bf 433}(7023), 289--291.

\bibitem[{Kimura} \& {Mann}(1998)]{kimura:1998}
{Kimura}, H. \& {Mann}, I., 1998.
\newblock {The Electric Charging of Interstellar Dust in the Solar System and
  Consequences for Its Dynamics}, {\it ApJ\/}, {\bf 499}(1), 454--462.

\bibitem[{Kleimann} et~al.(2022){Kleimann}, {Dialynas}, {Fraternale}, {Galli},
  {Heerikhuisen}, {Izmodenov}, {Kornbleuth}, {Opher}, \&
  {Pogorelov}]{kleimann_ea:2022}
{Kleimann}, J., {Dialynas}, K., {Fraternale}, F., {Galli}, A., {Heerikhuisen},
  J., {Izmodenov}, V., {Kornbleuth}, M., {Opher}, M., \& {Pogorelov}, N., 2022.
\newblock {The Structure of the Large-Scale Heliosphere as Seen by Current
  Models}, {\it \ssr\/}, {\bf 218}(4), 36.

\bibitem[{Kobayashi} et~al.(2020){Kobayashi}, {Karakas}, \&
  {Lugaro}]{kobayashi:2020}
{Kobayashi}, C., {Karakas}, A.~I., \& {Lugaro}, M., 2020.
\newblock {The Origin of Elements from Carbon to Uranium}, {\it \apj\/}, {\bf
  900}(2), 179.

\bibitem[{Koll} et~al.(2019){Koll}, {Korschinek}, {Faestermann},
  {G{\'o}mez-Guzm{\'a}n}, {Kipfstuhl}, {Merchel}, \& {Welch}]{Koll:2019}
{Koll}, D., {Korschinek}, G., {Faestermann}, T., {G{\'o}mez-Guzm{\'a}n}, J.~M.,
  {Kipfstuhl}, S., {Merchel}, S., \& {Welch}, J.~M., 2019.
\newblock {Interstellar $^{60}$Fe in Antarctica}, {\it \prl\/}, {\bf 123}(7),
  072701.

\bibitem[{Kr{\"u}ger} et~al.(2007){Kr{\"u}ger}, {Landgraf}, {Altobelli}, \&
  {Gr{\"u}n}]{krueger:2007}
{Kr{\"u}ger}, H., {Landgraf}, M., {Altobelli}, N., \& {Gr{\"u}n}, E., 2007.
\newblock {Interstellar Dust in the Solar System}, {\it Space Science
  Reviews\/}, {\bf 130}, 401--408.

\bibitem[{Kr{\"u}ger} et~al.(2015){Kr{\"u}ger}, {Strub}, {Gr{\"u}n}, \&
  {Sterken}]{krueger:2015}
{Kr{\"u}ger}, H., {Strub}, P., {Gr{\"u}n}, E., \& {Sterken}, V.~J., 2015.
\newblock {Sixteen Years of Ulysses Interstellar Dust Measurements in the Solar
  System. I. Mass Distribution and Gas-to-dust Mass Ratio}, {\it Astrophys.
  J.\/}, {\bf 812}, 139.

\bibitem[Kr{\"u}ger et~al.(2019)Kr{\"u}ger, Strub, Srama, Kobayashi, Arai,
  Kimura, Hirai, {Moragas-Klostermeyer}, Altobelli, Sterken, Agarwal, Sommer,
  \& Gr{\"u}n]{kruger:2019}
Kr{\"u}ger, H., Strub, P., Srama, R., Kobayashi, M., Arai, T., Kimura, H.,
  Hirai, T., {Moragas-Klostermeyer}, G., Altobelli, N., Sterken, V.~J.,
  Agarwal, J., Sommer, M., \& Gr{\"u}n, E., 2019.
\newblock Modelling {{DESTINY}}+ interplanetary and interstellar dust
  measurements en route to the active asteroid (3200) {{Phaethon}}, {\it
  Planetary and Space Science\/}, {\bf 172}, 22--42.

\bibitem[{Kr{\"u}ger} et~al.(2020){Kr{\"u}ger}, {Strub}, {Sommer}, {Altobelli},
  {Kimura}, {Lohse}, {Gr{\"u}n}, \& {Srama}]{krueger:2020}
{Kr{\"u}ger}, H., {Strub}, P., {Sommer}, M., {Altobelli}, N., {Kimura}, H.,
  {Lohse}, A.-K., {Gr{\"u}n}, E., \& {Srama}, R., 2020.
\newblock {Helios spacecraft data revisited: detection of cometary meteoroid
  trails by following in situ dust impacts}, {\it \aap\/}, {\bf 643}, A96.

\bibitem[{Landgraf} et~al.(2003){Landgraf}, {Kr{\"u}ger}, {Altobelli}, \&
  {Gr{\"u}n}]{landgraf:2003}
{Landgraf}, M., {Kr{\"u}ger}, H., {Altobelli}, N., \& {Gr{\"u}n}, E., 2003.
\newblock {Penetration of the heliosphere by the interstellar dust stream
  during solar maximum}, {\it Journal of Geophysical Research (Space
  Physics)\/}, {\bf 108}, 8030.

\bibitem[{Li}(2020)]{Li:2020}
{Li}, A., 2020.
\newblock {Spitzer's perspective of polycyclic aromatic hydrocarbons in
  galaxies}, {\it Nature Astronomy\/}, {\bf 4}, 339--351.

\bibitem[Li et~al.(2014)Li, Srama, Henkel, Sternovsky, Kempf, Wu, \&
  Gr{\"u}n]{li:2014}
Li, Y., Srama, R., Henkel, H., Sternovsky, Z., Kempf, S., Wu, Y., \& Gr{\"u}n,
  E., 2014.
\newblock Instrument study of the {{Lunar Dust eXplorer}} ({{LDX}}) for a lunar
  lander mission, {\it Advances in Space Research\/}, {\bf 54}(10), 2094--2100.

\bibitem[Li et~al.(2015)Li, Strack, Bugiel, Wu, \& Srama]{li:2015}
Li, Y., Strack, H., Bugiel, S., Wu, Y., \& Srama, R., 2015.
\newblock Instrument study of the {{Lunar Dust eXplorer}} ({{LDX}}) for a lunar
  lander mission {{II}}: {{Laboratory}} model calibration, {\it Advances in
  Space Research\/}, {\bf 56}(8), 1777--1783.

\bibitem[Li et~al.(2017)Li, Kempf, Simolka, Strack, Gr{\"u}n, \&
  Srama]{li:2017}
Li, Y., Kempf, S., Simolka, J., Strack, H., Gr{\"u}n, E., \& Srama, R., 2017.
\newblock Instrument concept of a single channel dust trajectory detector, {\it
  Advances in Space Research\/}, {\bf 59}(6), 1636--1641.

\bibitem[{Linde} \& {Gombosi}(2000)]{linde:2000}
{Linde}, T.~J. \& {Gombosi}, T.~I., 2000.
\newblock {Interstellar dust filtration at the heliospheric interface}, {\it
  \jgr\/}, {\bf 105}(A5), 10411--10418.

\bibitem[{Linsky} et~al.(2022){Linsky}, {Redfield}, {Ryder}, \&
  {Moebius}]{linsky:2022}
{Linsky}, J., {Redfield}, S., {Ryder}, D., \& {Moebius}, E., 2022.
\newblock {Inhomogeneity in the Local ISM and Its Relation to the Heliosphere},
  {\it \ssr\/}, {\bf 218}(3), 16.

\bibitem[{Love} \& {Brownlee}(1991)]{love:1991}
{Love}, S.~G. \& {Brownlee}, D.~E., 1991.
\newblock {Heating and thermal transformation of micrometeoroids entering the
  Earth's atmosphere}, {\it \icarus\/}, {\bf 89}(1), 26--43.

\bibitem[{Ma} et~al.(2013){Ma}, {Matthews}, {Land}, \& {Hyde}]{ma:2013}
{Ma}, Q., {Matthews}, L.~S., {Land}, V., \& {Hyde}, T.~W., 2013.
\newblock {Charging of Aggregate Grains in Astrophysical Environments}, {\it
  The Astrophysical Journal\/}, {\bf 763}, 77.

\bibitem[{Malaspina} \& {Wilson}(2016)]{malaspina:2016}
{Malaspina}, D.~M. \& {Wilson}, L.~B., 2016.
\newblock {A database of interplanetary and interstellar dust detected by the
  Wind spacecraft}, {\it Journal of Geophysical Research (Space Physics)\/},
  {\bf 121}(10), 9369--9377.

\bibitem[{Malaspina} et~al.(2014){Malaspina}, {Hor{\'a}nyi}, {Zaslavsky},
  {Goetz}, {Wilson}, \& {Kersten}]{malaspina:2014}
{Malaspina}, D.~M., {Hor{\'a}nyi}, M., {Zaslavsky}, A., {Goetz}, K., {Wilson},
  L.~B., \& {Kersten}, K., 2014.
\newblock {Interplanetary and interstellar dust observed by the Wind/WAVES
  electric field instrument}, {\it \grl\/}, {\bf 41}(2), 266--272.

\bibitem[{Mathis} et~al.(1977){Mathis}, {Rumpl}, \& {Nordsieck}]{mathis:1977}
{Mathis}, J.~S., {Rumpl}, W., \& {Nordsieck}, K.~H., 1977.
\newblock {The size distribution of interstellar grains}, {\it The
  Astrophysical Journal\/}, {\bf 217}, 425--433.

\bibitem[{McComas} et~al.(2008){McComas}, {Allegrini}, {Bagenal}, {Casey},
  {Delamere}, {Demkee}, {Dunn}, {Elliott}, {Hanley}, {Johnson}, {Langle},
  {Miller}, {Pope}, {Reno}, {Rodriguez}, {Schwadron}, {Valek}, \&
  {Weidner}]{McComas_ea:2008}
{McComas}, D., {Allegrini}, F., {Bagenal}, F., {Casey}, P., {Delamere}, P.,
  {Demkee}, D., {Dunn}, G., {Elliott}, H., {Hanley}, J., {Johnson}, K.,
  {Langle}, J., {Miller}, G., {Pope}, S., {Reno}, M., {Rodriguez}, B.,
  {Schwadron}, N., {Valek}, P., \& {Weidner}, S., 2008.
\newblock {The Solar Wind Around Pluto (SWAP) Instrument Aboard New Horizons},
  {\it \ssr\/}, {\bf 140}(1-4), 261--313.

\bibitem[{McComas} et~al.(2018){McComas}, {Christian}, {Schwadron}, {Fox},
  {Westlake}, {Allegrini}, {Baker}, {Biesecker}, {Bzowski}, {Clark}, {Cohen},
  {Cohen}, {Dayeh}, {Decker}, {de Nolfo}, {Desai}, {Ebert}, {Elliott}, {Fahr},
  {Frisch}, {Funsten}, {Fuselier}, {Galli}, {Galvin}, {Giacalone},
  {Gkioulidou}, {Guo}, {Horanyi}, {Isenberg}, {Janzen}, {Kistler}, {Korreck},
  {Kubiak}, {Kucharek}, {Larsen}, {Leske}, {Lugaz}, {Luhmann}, {Matthaeus},
  {Mitchell}, {Moebius}, {Ogasawara}, {Reisenfeld}, {Richardson}, {Russell},
  {Sok{\'o}{\l}}, {Spence}, {Skoug}, {Sternovsky}, {Swaczyna}, {Szalay},
  {Tokumaru}, {Wiedenbeck}, {Wurz}, {Zank}, \& {Zirnstein}]{mccomas2018}
{McComas}, D.~J., {Christian}, E.~R., {Schwadron}, N.~A., {Fox}, N.,
  {Westlake}, J., {Allegrini}, F., {Baker}, D.~N., {Biesecker}, D., {Bzowski},
  M., {Clark}, G., {Cohen}, C.~M.~S., {Cohen}, I., {Dayeh}, M.~A., {Decker},
  R., {de Nolfo}, G.~A., {Desai}, M.~I., {Ebert}, R.~W., {Elliott}, H.~A.,
  {Fahr}, H., {Frisch}, P.~C., {Funsten}, H.~O., {Fuselier}, S.~A., {Galli},
  A., {Galvin}, A.~B., {Giacalone}, J., {Gkioulidou}, M., {Guo}, F., {Horanyi},
  M., {Isenberg}, P., {Janzen}, P., {Kistler}, L.~M., {Korreck}, K., {Kubiak},
  M.~A., {Kucharek}, H., {Larsen}, B.~A., {Leske}, R.~A., {Lugaz}, N.,
  {Luhmann}, J., {Matthaeus}, W., {Mitchell}, D., {Moebius}, E., {Ogasawara},
  K., {Reisenfeld}, D.~B., {Richardson}, J.~D., {Russell}, C.~T.,
  {Sok{\'o}{\l}}, J.~M., {Spence}, H.~E., {Skoug}, R., {Sternovsky}, Z.,
  {Swaczyna}, P., {Szalay}, J.~R., {Tokumaru}, M., {Wiedenbeck}, M.~E., {Wurz},
  P., {Zank}, G.~P., \& {Zirnstein}, E.~J., 2018.
\newblock {Interstellar Mapping and Acceleration Probe (IMAP): A New NASA
  Mission}, {\it \ssr\/}, {\bf 214}(8), 116.

\bibitem[{McComas} et~al.(2021){McComas}, {Swaczyna}, {Szalay}, {Zirnstein},
  {Rankin}, {Elliott}, {Singer}, {Spencer}, {Stern}, \&
  {Weaver}]{McComas_ea:2021}
{McComas}, D.~J., {Swaczyna}, P., {Szalay}, J.~R., {Zirnstein}, E.~J.,
  {Rankin}, J.~S., {Elliott}, H.~A., {Singer}, K., {Spencer}, J., {Stern},
  S.~A., \& {Weaver}, H., 2021.
\newblock {Interstellar Pickup Ion Observations Halfway to the Termination
  Shock}, {\it \apjs\/}, {\bf 254}(1), 19.

\bibitem[{McNutt} et~al.(2008){McNutt}, {Livi}, {Gurnee}, {Hill}, {Cooper},
  {Andrews}, {Keath}, {Krimigis}, {Mitchell}, {Tossman}, {Bagenal}, {Boldt},
  {Bradley}, {Devereux}, {Ho}, {Jaskulek}, {Lefevere}, {Malcom}, {Marcus},
  {Hayes}, {Moore}, {Perry}, {Williams}, {Wilson}, {Brown}, {Kusterer}, \&
  {Vandegriff}]{mcnutt_ea:2008}
{McNutt}, R.~L., {Livi}, S.~A., {Gurnee}, R.~S., {Hill}, M.~E., {Cooper},
  K.~A., {Andrews}, G.~B., {Keath}, E.~P., {Krimigis}, S.~M., {Mitchell},
  D.~G., {Tossman}, B., {Bagenal}, F., {Boldt}, J.~D., {Bradley}, W.,
  {Devereux}, W.~S., {Ho}, G.~C., {Jaskulek}, S.~E., {Lefevere}, T.~W.,
  {Malcom}, H., {Marcus}, G.~A., {Hayes}, J.~R., {Moore}, G.~T., {Perry},
  M.~E., {Williams}, B.~D., {Wilson}, P., {Brown}, L.~E., {Kusterer}, M.~B., \&
  {Vandegriff}, J.~D., 2008.
\newblock {The Pluto Energetic Particle Spectrometer Science Investigation
  (PEPSSI) on the New Horizons Mission}, {\it \ssr\/}, {\bf 140}(1-4),
  315--385.

\bibitem[McNutt et~al.(2021)McNutt, Paul, Brandt, \& Kinnison]{mcnutt:2021}
McNutt, R.~L., Paul, M.~V., Brandt, P.~C., \& Kinnison, J.~D., 2021.
\newblock {Interstellar Probe -- Humanity's Journey to Interstellar Space},
  {\it NASA Solar and Space Physics Mission Concept Study for the Solar and
  Space Physics 2023–2032 Decadal Survey\/}.

\bibitem[{McNutt} et~al.(2022){McNutt}, {Wimmer-Schweingruber}, {Gruntman},
  {Krimigis}, {Roelof}, {Brandt}, {Vernon}, {Paul}, {Stough}, \&
  {Kinnison}]{mcnutt:2022}
{McNutt}, R.~L., {Wimmer-Schweingruber}, R.~F., {Gruntman}, M., {Krimigis},
  S.~M., {Roelof}, E.~C., {Brandt}, P.~C., {Vernon}, S.~R., {Paul}, M.~V.,
  {Stough}, R.~W., \& {Kinnison}, J.~D., 2022.
\newblock {Interstellar probe - Destination: Universe!}, {\it Acta
  Astronautica\/}, {\bf 196}, 13--28.

\bibitem[{Miller} et~al.(2022){Miller}, {Fields}, {Chen}, {Ellis}, {Ertel},
  {Manweiler}, {Opher}, {Provornikova}, {Slavin}, {Sok{\'o}{\l}}, {Sterken},
  {Surman}, \& {Wang}]{Miller:2022}
{Miller}, J.~A., {Fields}, B.~D., {Chen}, T.~Y., {Ellis}, J., {Ertel}, A.~F.,
  {Manweiler}, J.~W., {Opher}, M., {Provornikova}, E., {Slavin}, J.~D.,
  {Sok{\'o}{\l}}, J., {Sterken}, V., {Surman}, R., \& {Wang}, X., 2022.
\newblock {Near-Earth Supernovae in the Past 10 Myr: Implications for the
  Heliosphere}, {\it arXiv e-prints\/}, p. arXiv:2209.03497.

\bibitem[Napoleoni et~al.(2023)Napoleoni, Klenner, Khawaja, Hillier, \&
  Postberg]{napoleoni:2023}
Napoleoni, M., Klenner, F., Khawaja, N., Hillier, J.~K., \& Postberg, F., 2023.
\newblock Mass {{Spectrometric Fingerprints}} of {{Organic Compounds}} in
  {{NaCl-Rich Ice Grains}} from {{Europa}} and {{Enceladus}}, {\it ACS Earth
  and Space Chemistry\/}.

\bibitem[{Nesvorn{\'y}} et~al.(2010){Nesvorn{\'y}}, {Jenniskens}, {Levison},
  {Bottke}, {Vokrouhlick{\'y}}, \& {Gounelle}]{nesvorny:2010}
{Nesvorn{\'y}}, D., {Jenniskens}, P., {Levison}, H.~F., {Bottke}, W.~F.,
  {Vokrouhlick{\'y}}, D., \& {Gounelle}, M., 2010.
\newblock {Cometary Origin of the Zodiacal Cloud and Carbonaceous
  Micrometeorites. Implications for Hot Debris Disks}, {\it \apj\/}, {\bf
  713}(2), 816--836.

\bibitem[{Opher} \& {Loeb}(2022)]{opher2022a}
{Opher}, M. \& {Loeb}, A., 2022.
\newblock {Terrestrial Impact from the Passage of the Solar System through a
  Cold Cloud a Few Million Years Ago}, {\it arXiv e-prints\/}, p.
  arXiv:2202.01813.

\bibitem[Piquette et~al.(2019)Piquette, Poppe, Bernardoni, Szalay, James,
  Hor{\'a}nyi, Stern, Weaver, Spencer, \& Olkin]{piquette:2019}
Piquette, M., Poppe, A.~R., Bernardoni, E., Szalay, J.~R., James, D.,
  Hor{\'a}nyi, M., Stern, S.~A., Weaver, H., Spencer, J., \& Olkin, C., 2019.
\newblock Student {{Dust Counter}}: {{Status}} report at 38 {{AU}}, {\it
  Icarus\/}, {\bf 321}, 116--125.

\bibitem[{Poppe} et~al.(2022){Poppe}, {Horanyi}, {Szalay}, {Sternovsky},
  {Kempf}, {Bouwman}, {Postberg}, \& {Srama}]{poppe:2022}
{Poppe}, A., {Horanyi}, M., {Szalay}, J., {Sternovsky}, Z., {Kempf}, S.,
  {Bouwman}, J., {Postberg}, F., \& {Srama}, R., 2022.
\newblock {The interactions of Interstellar Dust with our Heliosphere (White
  paper submitted to for the decadal survey in Solar and Space Physics
  (Heliophysics) 2024-2033)}, {\it Bulletin of the AAS (BAAS)\/}.

\bibitem[{Poppe}(2016)]{poppe:2016}
{Poppe}, A.~R., 2016.
\newblock {An improved model for interplanetary dust fluxes in the outer Solar
  System}, {\it \icarus\/}, {\bf 264}, 369--386.

\bibitem[{Posner} et~al.(2021){Posner}, {Arge}, {Staub}, {StCyr}, {Folta},
  {Solanki}, {Strauss}, {Effenberger}, {Gandorfer}, {Heber}, {Henney},
  {Hirzberger}, {Jones}, {K{\"u}hl}, {Malandraki}, \& {Sterken}]{posner:2021}
{Posner}, A., {Arge}, C.~N., {Staub}, J., {StCyr}, O.~C., {Folta}, D.,
  {Solanki}, S.~K., {Strauss}, R.~D.~T., {Effenberger}, F., {Gandorfer}, A.,
  {Heber}, B., {Henney}, C.~J., {Hirzberger}, J., {Jones}, S.~I., {K{\"u}hl},
  P., {Malandraki}, O., \& {Sterken}, V.~J., 2021.
\newblock {A Multi-Purpose Heliophysics L4 Mission}, {\it Space Weather\/},
  {\bf 19}(9), e02777.

\bibitem[{Pusack} et~al.(2021){Pusack}, {Malaspina}, {Szalay}, {Bale}, {Goetz},
  {MacDowall}, \& {Pulupa}]{pusack:2021}
{Pusack}, A., {Malaspina}, D.~M., {Szalay}, J.~R., {Bale}, S.~D., {Goetz}, K.,
  {MacDowall}, R.~J., \& {Pulupa}, M., 2021.
\newblock {Dust Directionality and an Anomalous Interplanetary Dust Population
  Detected by the Parker Solar Probe}, {\it The Planetary Science Journal\/},
  {\bf 2}(5), 186.

\bibitem[{Richardson} et~al.(2008){Richardson}, {Kasper}, {Wang}, {Belcher}, \&
  {Lazarus}]{richardson_ea:2008}
{Richardson}, J.~D., {Kasper}, J.~C., {Wang}, C., {Belcher}, J.~W., \&
  {Lazarus}, A.~J., 2008.
\newblock {Cool heliosheath plasma and deceleration of the upstream solar wind
  at the termination shock}, {\it \nat\/}, {\bf 454}(7200), 63--66.

\bibitem[{Rietmeijer}(1996)]{rietmeijer:1996}
{Rietmeijer}, F. J.~M., 1996.
\newblock {CM-like interplanetary dust particles in lower stratosphere during
  1989 October and 1991 June/July}, {\it Meteorit.\ Planet.\ Sci.\/}, {\bf
  31}(2), 278--288.

\bibitem[{Schwadron} \& {Gloeckler}(2007)]{schwadron+2007}
{Schwadron}, N.~A. \& {Gloeckler}, G., 2007.
\newblock {Pickup Ions and Cosmic Rays from Dust in the Heliosphere}, {\it
  \ssr\/}, {\bf 130}(1-4), 283--291.

\bibitem[{Schwadron} et~al.(2002{\natexlab{a}}){Schwadron}, {Combi}, {Huebner},
  \& {McComas}]{schwadron+2002}
{Schwadron}, N.~A., {Combi}, M., {Huebner}, W., \& {McComas}, D.~J.,
  2002{\natexlab{a}}.
\newblock {The outer source of pickup ions and anomalous cosmic rays}, {\it
  \grl\/}, {\bf 29}(20), 1993.

\bibitem[{Schwadron} et~al.(2002{\natexlab{b}}){Schwadron}, {Combi}, {Huebner},
  \& {McComas}]{schwadron2002+}
{Schwadron}, N.~A., {Combi}, M., {Huebner}, W., \& {McComas}, D.~J.,
  2002{\natexlab{b}}.
\newblock {The outer source of pickup ions and anomalous cosmic rays}, {\it
  \grl\/}, {\bf 29}(20), 1993.

\bibitem[{Shen} et~al.(2021){Shen}, {Sternovsky}, {Hor{\'a}nyi}, {Hsu}, \&
  {Malaspina}]{shen:2021}
{Shen}, M.~M., {Sternovsky}, Z., {Hor{\'a}nyi}, M., {Hsu}, H.-W., \&
  {Malaspina}, D.~M., 2021.
\newblock {Laboratory Study of Antenna Signals Generated by Dust Impacts on
  Spacecraft}, {\it Journal of Geophysical Research (Space Physics)\/}, {\bf
  126}(4), e28965.

\bibitem[{Shen} et~al.(2023){Shen}, {Sternovsky}, \& {Malaspina}]{shen:2023}
{Shen}, M.~M., {Sternovsky}, Z., \& {Malaspina}, D.~M., 2023.
\newblock {Variability of Antenna Signals From Dust Impacts}, {\it Journal of
  Geophysical Research (Space Physics)\/}, {\bf 128}(4), e2022JA030981.

\bibitem[{Simolka} et~al.(2022){Simolka}, {Bauer}, {Exle}, {Fr{\"o}hlich},
  {Gl{\"a}ser}, {Ingerl}, {Li}, {Sommer}, {Strack}, {Henkel}, {Wagner}, \&
  {Srama}]{simolka:2022}
{Simolka}, J., {Bauer}, M., {Exle}, A., {Fr{\"o}hlich}, P., {Gl{\"a}ser}, J.,
  {Ingerl}, S., {Li}, Y., {Sommer}, M., {Strack}, H., {Henkel}, H., {Wagner},
  C., \& {Srama}, R., 2022.
\newblock {Development of the Destiny+ Dust Telescope}, in {\em European
  Planetary Science Congress\/}, pp. EPSC2022--1070.

\bibitem[{Simpson} \& {Tuzzolino}(1985)]{simpson:1985}
{Simpson}, J.~A. \& {Tuzzolino}, A.~J., 1985.
\newblock {Polarized polymer films as electronic pulse detectors of cosmic dust
  particles}, {\it Nuclear Instruments and Methods in Physics Research A\/},
  {\bf 236}(1), 187--202.

\bibitem[{Slavin} et~al.(2012){Slavin}, {Frisch}, {M{\"u}ller}, {Heerikhuisen},
  {Pogorelov}, {Reach}, \& {Zank}]{slavin:2012}
{Slavin}, J.~D., {Frisch}, P.~C., {M{\"u}ller}, H.-R., {Heerikhuisen}, J.,
  {Pogorelov}, N.~V., {Reach}, W.~T., \& {Zank}, G., 2012.
\newblock {Trajectories and Distribution of Interstellar Dust Grains in the
  Heliosphere}, {\it ApJ\/}, {\bf 760}(1), 46.

\bibitem[{Soja} et~al.(2014){Soja}, {Sommer}, {Herzog}, {Srama}, {Gr{\"u}n},
  {Rodmann}, {Strub}, {Vaubaillon}, {Hornig}, \& {Bausch}]{soja:2014}
{Soja}, R.~H., {Sommer}, M., {Herzog}, J., {Srama}, R., {Gr{\"u}n}, E.,
  {Rodmann}, J., {Strub}, P., {Vaubaillon}, J., {Hornig}, A., \& {Bausch}, L.,
  2014.
\newblock {The Interplanetary Meteoroid Environment for eXploration - (IMEX)
  project}, in {\em Proceedings of the International Meteor Conference, Giron,
  France, 18-21 September 2014\/}, pp. 146--149.

\bibitem[Srama et~al.(2004{\natexlab{a}})Srama, Ahrens, Altobelli, Auer,
  Bradley, Burton, Dikarev, Economou, Fechtig, G{\"o}rlich, Grande, Graps,
  Gr{\"u}n, Havnes, Helfert, Horanyi, Igenbergs, Jessberger, Johnson, Kempf,
  Krivov, Kr{\"u}ger, {Mocker-Ahlreep}, {Moragas-Klostermeyer}, Lamy, Landgraf,
  Linkert, Linkert, Lura, McDonnell, M{\"o}hlmann, Morfill, M{\"u}ller, Roy,
  Sch{\"a}fer, Schlotzhauer, Schwehm, Spahn, St{\"u}big, Svestka,
  Tschernjawski, Tuzzolino, W{\"a}sch, \& Zook]{srama:2004}
Srama, R., Ahrens, T.~J., Altobelli, N., Auer, S., Bradley, J.~G., Burton, M.,
  Dikarev, V.~V., Economou, T., Fechtig, H., G{\"o}rlich, M., Grande, M.,
  Graps, A., Gr{\"u}n, E., Havnes, O., Helfert, S., Horanyi, M., Igenbergs, E.,
  Jessberger, E.~K., Johnson, T.~V., Kempf, S., Krivov, A.~V., Kr{\"u}ger, H.,
  {Mocker-Ahlreep}, A., {Moragas-Klostermeyer}, G., Lamy, P., Landgraf, M.,
  Linkert, D., Linkert, G., Lura, F., McDonnell, J. A.~M., M{\"o}hlmann, D.,
  Morfill, G.~E., M{\"u}ller, M., Roy, M., Sch{\"a}fer, G., Schlotzhauer, G.,
  Schwehm, G.~H., Spahn, F., St{\"u}big, M., Svestka, J., Tschernjawski, V.,
  Tuzzolino, A.~J., W{\"a}sch, R., \& Zook, H.~A., 2004{\natexlab{a}}.
\newblock The {{Cassini Cosmic Dust Analyzer}}, {\it Space Science Reviews\/},
  {\bf 114}, 465--518.

\bibitem[Srama et~al.(2004{\natexlab{b}})Srama, Rachev, Srowig, Kempf,
  {Moragas-Klostermeyer}, Kr{\"u}ger, Auer, Glasmacher, \&
  R{\"u}nn]{srama:2004-2}
Srama, R., Rachev, M., Srowig, A., Kempf, S., {Moragas-Klostermeyer}, G.,
  Kr{\"u}ger, H., Auer, S., Glasmacher, A., \& R{\"u}nn, E.,
  2004{\natexlab{b}}.
\newblock Dust {{Astronomy}} with a {{Dust Telescope}}, in {\em Proceedings of
  the 37th {{ESLAB Symposium}}\/}, vol. ESA SP-543, pp. 73--78.

\bibitem[{Srama} et~al.(2007){Srama}, {Kempf}, {Moragas-Klostermeyer},
  {Landgraf}, {Helfert}, {Sternovsky}, {Rachev}, \& {Gruen}]{srama:2007}
{Srama}, R., {Kempf}, S., {Moragas-Klostermeyer}, G., {Landgraf}, M.,
  {Helfert}, S., {Sternovsky}, Z., {Rachev}, M., \& {Gruen}, E., 2007.
\newblock {Laboratory Tests of the Large Area Mass Analyser}, in {\em Dust in
  Planetary Systems\/}, vol. 643 of {\bf ESA SP}, pp. 209--212.

\bibitem[Srama et~al.(2009)Srama, Woiwode, Postberg, Armes, Fujii, Dupin,
  {Ormond-Prout}, Sternovsky, Kempf, {Moragas-Klostermeyer}, Mocker, \&
  Gr{\"u}n]{srama:2009}
Srama, R., Woiwode, W., Postberg, F., Armes, S.~P., Fujii, S., Dupin, D.,
  {Ormond-Prout}, J., Sternovsky, Z., Kempf, S., {Moragas-Klostermeyer}, G.,
  Mocker, A., \& Gr{\"u}n, E., 2009.
\newblock Mass spectrometry of hyper-velocity impacts of organic micrograins,
  {\it Rapid Communications in Mass Spectrometry\/}, {\bf 23}(24), 3895--3906.

\bibitem[{Stephan}(2001)]{stephan:2001}
{Stephan}, T., 2001.
\newblock {TOF-SIMS in cosmochemistry}, {\it \planss\/}, {\bf 49}(9), 859--906.

\bibitem[{Sterken}(2023{\natexlab{a}})]{sterken:2023a}
{Sterken}, V.~J., 2023{\natexlab{a}}.
\newblock {Dust measurement opportunities from the lunar gateway: a science
  based approach}, {\it In prep.\/}.

\bibitem[{Sterken}(2023{\natexlab{b}})]{sterken:2023b}
{Sterken}, V.~J., 2023{\natexlab{b}}.
\newblock {The DOLPHIN mission concept}, {\it In prep.\/}.

\bibitem[{Sterken} et~al.(2012){Sterken}, {Altobelli}, {Kempf}, {Schwehm},
  {Srama}, \& {Gr{\"u}n}]{sterken:2012}
{Sterken}, V.~J., {Altobelli}, N., {Kempf}, S., {Schwehm}, G., {Srama}, R., \&
  {Gr{\"u}n}, E., 2012.
\newblock {The flow of interstellar dust into the solar system}, {\it A\&A\/},
  {\bf 538}, A102.

\bibitem[{Sterken} et~al.(2015){Sterken}, {Strub}, {Kr{\"u}ger}, {von Steiger},
  \& {Frisch}]{sterken:2015}
{Sterken}, V.~J., {Strub}, P., {Kr{\"u}ger}, H., {von Steiger}, R., \&
  {Frisch}, P., 2015.
\newblock {Sixteen Years of Ulysses Interstellar Dust Measurements in the Solar
  System. III. Simulations and Data Unveil New Insights into Local Interstellar
  Dust}, {\it Astrophys. J.\/}, {\bf 812}, 141.

\bibitem[{Sterken} et~al.(2019){Sterken}, {Westphal}, {Altobelli}, {Malaspina},
  \& {Postberg}]{sterken:2019}
{Sterken}, V.~J., {Westphal}, A.~J., {Altobelli}, N., {Malaspina}, D., \&
  {Postberg}, F., 2019.
\newblock {Interstellar Dust in the Solar System}, {\it Space~Sci.~Rev.\/},
  {\bf 215}(7), 43.

\bibitem[{Sterken} et~al.(2022){Sterken}, {Hunziker}, {Dialynas}, {Herbst},
  {Li}, {Baalmann}, \& {Scherer}]{sterken:2022}
{Sterken}, V.~J., {Hunziker}, S., {Dialynas}, K., {Herbst}, K., {Li}, A.,
  {Baalmann}, L.~R., \& {Scherer}, K., 2022.
\newblock {Synergies between interstellar dust and heliospheric science with an
  Interstellar Probe (White paper submitted to for the decadal survey in Solar
  and Space Physics (Heliophysics) 2024-2033)}, {\it Bulletin of the AAS
  (BAAS)\/}.

\bibitem[{Sternovsky} et~al.(2007){Sternovsky}, {Amyx}, {Bano}, {Landgraf},
  {Horanyi}, {Knappmiller}, {Robertson}, {Gr{\"u}n}, {Srama}, \&
  {Auer}]{sternovsky:2007}
{Sternovsky}, Z., {Amyx}, K., {Bano}, G., {Landgraf}, M., {Horanyi}, M.,
  {Knappmiller}, S., {Robertson}, S., {Gr{\"u}n}, E., {Srama}, R., \& {Auer},
  S., 2007.
\newblock {Large area mass analyzer instrument for the chemical analysis of
  interstellar dust particles}, {\it Review of Scientific Instruments\/}, {\bf
  78}(1), 014501--014501--10.

\bibitem[{Stone} et~al.(2005){Stone}, {Cummings}, {McDonald}, {Heikkila},
  {Lal}, \& {Webber}]{stone_ea:2005}
{Stone}, E.~C., {Cummings}, A.~C., {McDonald}, F.~B., {Heikkila}, B.~C., {Lal},
  N., \& {Webber}, W.~R., 2005.
\newblock {Voyager 1 Explores the Termination Shock Region and the Heliosheath
  Beyond}, {\it Science\/}, {\bf 309}(5743), 2017--2020.

\bibitem[{Stone} et~al.(2008){Stone}, {Cummings}, {McDonald}, {Heikkila},
  {Lal}, \& {Webber}]{stone_ea:2008}
{Stone}, E.~C., {Cummings}, A.~C., {McDonald}, F.~B., {Heikkila}, B.~C., {Lal},
  N., \& {Webber}, W.~R., 2008.
\newblock {An asymmetric solar wind termination shock}, {\it \nat\/}, {\bf
  454}(7200), 71--74.

\bibitem[{Strub} et~al.(2015){Strub}, {Kr{\"u}ger}, \& {Sterken}]{strub:2015}
{Strub}, P., {Kr{\"u}ger}, H., \& {Sterken}, V.~J., 2015.
\newblock {Sixteen Years of Ulysses Interstellar Dust Measurements in the Solar
  System. II. Fluctuations in the Dust Flow from the Data}, {\it Astrophys.
  J.\/}, {\bf 812}, 140.

\bibitem[{Strub} et~al.(2019){Strub}, {Sterken}, {Soja}, {Kr{\"u}ger},
  {Gr{\"u}n}, \& {Srama}]{strub:2019}
{Strub}, P., {Sterken}, V.~J., {Soja}, R., {Kr{\"u}ger}, H., {Gr{\"u}n}, E., \&
  {Srama}, R., 2019.
\newblock {Heliospheric modulation of the interstellar dust flow on to Earth},
  {\it A\&A\/}, {\bf 621}, A54.

\bibitem[{Swaczyna} et~al.(2022){Swaczyna}, {Schwadron}, {M{\"o}bius},
  {Bzowski}, {Frisch}, {Linsky}, {McComas}, {Rahmanifard}, {Redfield},
  {Winslow}, {Wood}, \& {Zank}]{swaczyna:2022}
{Swaczyna}, P., {Schwadron}, N.~A., {M{\"o}bius}, E., {Bzowski}, M., {Frisch},
  P.~C., {Linsky}, J.~L., {McComas}, D.~J., {Rahmanifard}, F., {Redfield}, S.,
  {Winslow}, R.~M., {Wood}, B.~E., \& {Zank}, G.~P., 2022.
\newblock {Mixing Interstellar Clouds Surrounding the Sun}, {\it \apjl\/}, {\bf
  937}(2), L32.

\bibitem[{Szalay} et~al.(2021){Szalay}, {Pokorn{\'y}}, {Malaspina}, {Pusack},
  {Bale}, {Battams}, {Gasque}, {Goetz}, {Kr{\"u}ger}, {McComas}, {Schwadron},
  \& {Strub}]{szalay+2021}
{Szalay}, J.~R., {Pokorn{\'y}}, P., {Malaspina}, D.~M., {Pusack}, A., {Bale},
  S.~D., {Battams}, K., {Gasque}, L.~C., {Goetz}, K., {Kr{\"u}ger}, H.,
  {McComas}, D.~J., {Schwadron}, N.~A., \& {Strub}, P., 2021.
\newblock {Collisional Evolution of the Inner Zodiacal Cloud}, {\it Planet.\
  Sci.\ J\/}, {\bf 2}(5), 185.

\bibitem[{Timmes} et~al.(1995){Timmes}, {Woosley}, \& {Weaver}]{timmes:1995}
{Timmes}, F.~X., {Woosley}, S.~E., \& {Weaver}, T.~A., 1995.
\newblock {Galactic Chemical Evolution: Hydrogen through Zinc}, {\it \apjs\/},
  {\bf 98}, 617.

\bibitem[{Utterback} \& {Kissel}(1990)]{utterback:1990}
{Utterback}, N.~G. \& {Kissel}, J., 1990.
\newblock {Attogram Dust Cloud a Million Kilometers from Comet Halley}, {\it
  \aj\/}, {\bf 100}, 1315.

\bibitem[{Wang} et~al.(2015){Wang}, {Li}, \& {Jiang}]{wang:2015}
{Wang}, S., {Li}, A., \& {Jiang}, B.~W., 2015.
\newblock {Very Large Interstellar Grains as Evidenced by the Mid-infrared
  Extinction}, {\it The Astrophysical Journal\/}, {\bf 811}, 38.

\bibitem[{Wehry} et~al.(2004){Wehry}, {Kr{\"u}ger}, \& {Gr{\"u}n}]{wehry:2004}
{Wehry}, A., {Kr{\"u}ger}, H., \& {Gr{\"u}n}, E., 2004.
\newblock {Analysis of Ulysses data: Radiation pressure effects on dust
  particles}, {\it \aap\/}, {\bf 419}, 1169--1174.

\bibitem[{Weingartner} \& {Draine}(2001)]{weingartner:2001}
{Weingartner}, J.~C. \& {Draine}, B.~T., 2001.
\newblock {Dust Grain-Size Distributions and Extinction in the Milky Way, Large
  Magellanic Cloud, and Small Magellanic Cloud}, {\it \apj\/}, {\bf 548}(1),
  296--309.

\bibitem[{Westphal} et~al.(2014){Westphal}, {Stroud}, {Bechtel}, {Brenker},
  {Butterworth}, {Flynn}, {Frank}, {Gainsforth}, {Hillier}, {Postberg},
  {Simionovici}, {Sterken}, {Nittler}, {Allen}, {Anderson}, {Ansari}, {Bajt},
  {Bastien}, {Bassim}, {Bridges}, {Brownlee}, {Burchell}, {Burghammer},
  {Changela}, {Cloetens}, {Davis}, {Doll}, {Floss}, {Gr{\"u}n}, {Heck},
  {Hoppe}, {Hudson}, {Huth}, {Kearsley}, {King}, {Lai}, {Leitner}, {Lemelle},
  {Leonard}, {Leroux}, {Lettieri}, {Marchant}, {Ogliore}, {Ong}, {Price},
  {Sandford}, {Tresseras}, {Schmitz}, {Schoonjans}, {Schreiber}, {Silversmit},
  {Sol{\'e}}, {Srama}, {Stadermann}, {Stephan}, {Stodolna}, {Sutton},
  {Trieloff}, {Tsou}, {Tyliszczak}, {Vekemans}, {Vincze}, {Von Korff},
  {Wordsworth}, {Zevin}, {Zolensky}, \& {aff14}]{westphal:2014}
{Westphal}, A.~J., {Stroud}, R.~M., {Bechtel}, H.~A., {Brenker}, F.~E.,
  {Butterworth}, A.~L., {Flynn}, G.~J., {Frank}, D.~R., {Gainsforth}, Z.,
  {Hillier}, J.~K., {Postberg}, F., {Simionovici}, A.~S., {Sterken}, V.~J.,
  {Nittler}, L.~R., {Allen}, C., {Anderson}, D., {Ansari}, A., {Bajt}, S.,
  {Bastien}, R.~K., {Bassim}, N., {Bridges}, J., {Brownlee}, D.~E., {Burchell},
  M., {Burghammer}, M., {Changela}, H., {Cloetens}, P., {Davis}, A.~M., {Doll},
  R., {Floss}, C., {Gr{\"u}n}, E., {Heck}, P.~R., {Hoppe}, P., {Hudson}, B.,
  {Huth}, J., {Kearsley}, A., {King}, A.~J., {Lai}, B., {Leitner}, J.,
  {Lemelle}, L., {Leonard}, A., {Leroux}, H., {Lettieri}, R., {Marchant}, W.,
  {Ogliore}, R., {Ong}, W.~J., {Price}, M.~C., {Sandford}, S.~A., {Tresseras},
  J.-A.~S., {Schmitz}, S., {Schoonjans}, T., {Schreiber}, K., {Silversmit}, G.,
  {Sol{\'e}}, V.~A., {Srama}, R., {Stadermann}, F., {Stephan}, T., {Stodolna},
  J., {Sutton}, S., {Trieloff}, M., {Tsou}, P., {Tyliszczak}, T., {Vekemans},
  B., {Vincze}, L., {Von Korff}, J., {Wordsworth}, N., {Zevin}, D., {Zolensky},
  M.~E., \& {aff14}, 2014.
\newblock {Evidence for interstellar origin of seven dust particles collected
  by the Stardust spacecraft}, {\it Science\/}, {\bf 345}(6198), 786--791.

\bibitem[{Wouterloot} et~al.(2008){Wouterloot}, {Henkel}, {Brand}, \&
  {Davis}]{wouterloot:2008}
{Wouterloot}, J.~G.~A., {Henkel}, C., {Brand}, J., \& {Davis}, G.~R., 2008.
\newblock {Galactic interstellar $^{18}$O/\{\^17\}O ratios - a radial
  gradient?}, {\it \aap\/}, {\bf 487}(1), 237--246.

\bibitem[{Wozniakiewicz} et~al.(2021){Wozniakiewicz}, {Bridges}, {Burchell},
  {Carey}, {Carpenter}, {Della Corte}, {Dignam}, {Genge}, {Hicks},
  {Hilchenbach}, {Hillier}, {Kearsley}, {Kr{\"u}ger}, {Merouane}, {Palomba},
  {Postberg}, {Schmidt}, {Srama}, {Trieloff}, {van-Ginneken}, \&
  {Sterken}]{wozniakiewicz:2021}
{Wozniakiewicz}, P.~J., {Bridges}, J., {Burchell}, M.~J., {Carey}, W.,
  {Carpenter}, J., {Della Corte}, V., {Dignam}, A., {Genge}, M.~J., {Hicks},
  L., {Hilchenbach}, M., {Hillier}, J., {Kearsley}, A.~T., {Kr{\"u}ger}, H.,
  {Merouane}, S., {Palomba}, E., {Postberg}, F., {Schmidt}, J., {Srama}, R.,
  {Trieloff}, M., {van-Ginneken}, M., \& {Sterken}, V.~J., 2021.
\newblock {A cosmic dust detection suite for the deep space Gateway}, {\it
  Advances in Space Research\/}, {\bf 68}(1), 85--104.

\bibitem[Xie et~al.(2011)Xie, Sternovsky, Gr{\"u}n, Auer, Duncan, Drake, Le,
  Horanyi, \& Srama]{xie:2011}
Xie, J., Sternovsky, Z., Gr{\"u}n, E., Auer, S., Duncan, N., Drake, K., Le, H.,
  Horanyi, M., \& Srama, R., 2011.
\newblock Dust trajectory sensor: {{Accuracy}} and data analysis, {\it Review
  of Scientific Instruments\/}, {\bf 82}(10), 105104.

\bibitem[{Zhukovska} et~al.(2016){Zhukovska}, {Dobbs}, {Jenkins}, \&
  {Klessen}]{zhukovska:2016}
{Zhukovska}, S., {Dobbs}, C., {Jenkins}, E.~B., \& {Klessen}, R.~S., 2016.
\newblock {Modeling Dust Evolution in Galaxies with a Multiphase, Inhomogeneous
  ISM}, {\it \apj\/}, {\bf 831}(2), 147.

\bibitem[{Zinner}(2014)]{zinner:2014}
{Zinner}, E., 2014.
\newblock {Presolar Grains}, in {\em Meteorites and Cosmochemical Processes\/},
  vol.~1, pp. 181--213, ed. {Davis}, A.~M.

\bibitem[{Zinner} et~al.(2006){Zinner}, {Nittler}, {Gallino}, {Karakas},
  {Lugaro}, {Straniero}, \& {Lattanzio}]{zinner:2006}
{Zinner}, E., {Nittler}, L.~R., {Gallino}, R., {Karakas}, A.~I., {Lugaro}, M.,
  {Straniero}, O., \& {Lattanzio}, J.~C., 2006.
\newblock {Silicon and Carbon Isotopic Ratios in AGB Stars: SiC Grain Data,
  Models, and the Galactic Evolution of the Si Isotopes}, {\it \apj\/}, {\bf
  650}(1), 350--373.

\bibitem[{Zirnstein} et~al.(2022){Zirnstein}, {M{\"o}bius}, {Zhang}, {Bower},
  {Elliott}, {McComas}, {Pogorelov}, \& {Swaczyna}]{zirnstein+2022}
{Zirnstein}, E.~J., {M{\"o}bius}, E., {Zhang}, M., {Bower}, J., {Elliott},
  H.~A., {McComas}, D.~J., {Pogorelov}, N.~V., \& {Swaczyna}, P., 2022.
\newblock {In Situ Observations of Interstellar Pickup Ions from 1 au to the
  Outer Heliosphere}, {\it \ssr\/}, {\bf 218}(4), 28.

\bibitem[{Zook} et~al.(1996){Zook}, {Grun}, {Baguhl}, {Hamilton}, {Linkert},
  {Liou}, {Forsyth}, \& {Phillips}]{zook:1996}
{Zook}, H.~A., {Grun}, E., {Baguhl}, M., {Hamilton}, D.~P., {Linkert}, G.,
  {Liou}, J.~C., {Forsyth}, R., \& {Phillips}, J.~L., 1996.
\newblock {Solar Wind Magnetic Field Bending of Jovian Dust Trajectories}, {\it
  Science\/}, {\bf 274}(5292), 1501--1503.

\bibitem[Zymak et~al.(2023)Zymak, {\v Z}abka, Pol{\'a}{\v s}ek, Sanderink,
  Lebreton, Gaubicher, Cherville, Zymakov{\'a}, \& Briois]{zymak:2023}
Zymak, I., {\v Z}abka, J., Pol{\'a}{\v s}ek, M., Sanderink, A., Lebreton,
  J.-P., Gaubicher, B., Cherville, B., Zymakov{\'a}, A., \& Briois, C., 2023.
\newblock A {{High-Resolution Mass Spectrometer}} for the {{Experimental
  Study}} of the {{Gas Composition}} in {{Planetary Environments}}: {{First
  Laboratory Results}}, {\it Aerospace\/}, {\bf 10}(6), 522.

\end{thebibliography}





\appendix
\section{Supporting missions and mission concepts}
\label{sec:supportmissions}

\begin{table*}
\begin{center}
\begin{tabular}{p{1.5cm}|p{1.9cm}p{1.6cm}p{1.7cm}p{8.8cm}} 
 \hline
Mission / \newline Concept   & Version & Period of\newline flight & Orbit & Primary science goals \\ \hline \hline

IMAP & \citet{mccomas2018} & 2025 \--- & Sun-Earth L$_1$ & Composition and properties of the LISM (incl. dust);\newline Dynamics/evolution of the heliosheath;\newline Interaction of solar wind magnetic field and interstellar magnetic field;\newline Particle injection and acceleration processes \\ \hline 

ISP & \citet{brandt:2022}; \newline \citet{isp:website} & 2036 \--- 2086+ & Escape trajectory & Global nature of the heliosphere: dynamics \& evolution, evolutionary history; \newline 
Properties of the ISM: gas, dust, magnetic field, low-energy cosmic rays; \newline Exploration of the large-scale circumsolar dust debris disk; \newline Flyby observations of Kuiper Belt objets and planetesimals; \newline Unobscured mapping of the Cosmic Infrared Background\\ \hline

DOLPHIN (2022) & \citet{sterken:2023b} & 2031 \--- 2035+ & incl.\ 23$^\circ$\newline at 1~AU & Interstellar dust: composition, abundances, size distribution and its modulation by the heliosphere, dust-heliosphere interaction and physics (incl. PUI), role of dust/PUI in heliosphere-LISM pressure balance; \newline Structure and time-variable properties of the zodiacal dust cloud; \newline Composition of cometary dust\\ \hline

SunCHASER & \citet{posner:2021} & tbd & incl.\ 14$^\circ$ at Sun-Earth L$_4$, \newline potentially also at L$_5$ and L$_1$ & Solar energetic particle forecasting; \newline Improve model of inner heliosphere solar wind and magnetic field; \newline Long-term forecasting of solar activity; \newline Dust populations in near-Sun environment  \\ \hline

Lunar\newline Gateway & \citet{sterken:2023a, wozniakiewicz:2021} & tbd & lunar halo orbit & presence/absence of human-made debris in lunar orbit; \newline Composition and fluxes of asteroidal and cometary dust, separately; \newline Interstellar dust: composition, species abundances, size distribution, morphology, directionality; \newline Organic component of dust grains; \newline Shape, dynamics and physics of heliosphere\\ \hline
\end{tabular}
\caption{Summary of the missions and mission concepts considered in this publication with strong dust-heliosphere synergies.}\label{tab:missions_comp}
\end{center}
\end{table*}

\bsp	
\label{lastpage}
\end{document}